# Quantitative electronic structure and work-function changes of liquid water induced by solute

Bruno Credidio[1,2], Michele Pugini[1], Sebastian Malerz[1], Florian Trinter[1,3], Uwe Hergenhahn[1], Iain Wilkinson,[4] Stephan Thürmer[5]*, and Bernd Winter[1]*

[1] Molecular Physics Department, Fritz-Haber-Institut der Max-Planck-Gesellschaft, Faradayweg 4-6, 14195 Berlin, Germany
[2] Institute for Chemical Sciences and Engineering (ISIC), École Polytechnique Fédérale de Lausanne (EPFL), 1015 Lausanne, Switzerland
[3] Institut für Kernphysik, Goethe-Universität, Max-von-Laue-Straße 1, 60438 Frankfurt am Main, Germany
[4] Department of Locally-Sensitive & Time-Resolved Spectroscopy, Helmholtz-Zentrum Berlin für Materialien und Energie, Hahn-Meitner-Platz 1, 14109 Berlin, Germany
[5] Department of Chemistry, Graduate School of Science, Kyoto University, Kitashirakawa-Oiwakecho, Sakyo-Ku, Kyoto 606-8502, Japan

ORCID
BC: 0000-0003-0348-0778
MP: 0000-0003-2406-831X
SM: 0000-0001-9570-3494
FT: 0000-0002-0891-9180
UH: 0000-0003-3396-4511
IW: 0000-0001-9561-5056
ST: 0000-0002-8146-4573
BW: 0000-0002-5597-8888

*Corresponding authors: thuermer@kuchem.kyoto-u.ac.jp; winter@fhi-berlin.mpg.de

**Abstract**

Recent advancement in quantitative liquid-jet photoelectron spectroscopy enables the accurate determination of the absolute-scale electronic energetics of liquids and species in solution. The major objective of the present work is the determination of the absolute lowest-ionization energy of liquid water, corresponding to the $1b_1$ orbital electron liberation, which is found to vary upon solute addition, and depends on the solute concentration. We discuss two prototypical aqueous salt solutions, $NaI_{(aq)}$ and tetrabutylammonium iodide, $TBAI_{(aq)}$, with the latter being a strong surfactant. Our results reveal considerably different behavior of the liquid water $1b_1$ binding energy in each case. In the $NaI_{(aq)}$ solutions, the $1b_1$ energy increases by about 0.3 eV upon increasing the salt concentration from very dilute to near-saturation concentrations, whereas for TBAI the energy decreases by about 0.7 eV upon formation of a TBAI surface layer. The photoelectron spectra also allow us to quantify the solute-induced effects on the solute binding energies, as inferred from concentration-dependent energy shifts of the $I^-$ 5p binding energy. For $NaI_{(aq)}$, an almost identical $I^-$ 5p shift is found as for the water $1b_1$ binding energy, with a larger shift occurring in the opposite direction for the $TBAI_{(aq)}$ solution. We show that the evolution of the water $1b_1$ energy in the $NaI_{(aq)}$ solutions can be primarily assigned to a change of water's electronic structure in the solution bulk. In contrast, apparent changes of the $1b_1$ energy for $TBAI_{(aq)}$ solutions can be related to changes of the solution work function which could arise from surface molecular dipoles. Furthermore, for both of the solutions studied here, the measured water $1b_1$ binding energies can be correlated with the extensive solution molecular structure changes occurring at high salt concentrations, where in the case of $NaI_{(aq)}$, too few water molecules exist to hydrate individual ions and the solution adopts a crystalline-like phase. We also comment on the concentration-dependent shape of the second, $3a_1$ orbital liquid water ionization feature which is a sensitive signature of water–water hydrogen bond interactions.

## I. Introduction

Experimental access to absolute binding energies (BEs) from aqueous solutions has been a principal goal in liquid-jet photoelectron spectroscopy (LJ-PES) but can only now be accomplished thanks to a recent extension of the method's capabilities, by acquiring additional spectral information. In particular, here we make use of a novel energy referencing scheme, which has been described in detail in our recent publication[1] and is briefly summarized in the following. The key concept is to not only measure a desired photoelectron peak, *i.e.*, the respective kinetic energy (KE) associated with a given ionization feature, but to also measure the distribution of the spectral low-energy tail (LET) arising from various electron scattering processes,[2] and especially the energy of the cutoff feature, $E_{cut}$, of this scattering distribution. Briefly, this spectral cutoff indicates the lower bound of electron KEs *within* the liquid which can still overcome the surface barrier and be expelled from the solution. An electron imparted with an energy equivalent to the BE via photoabsorption will be found outside the solution with zero KE, under the proviso that electron does not undergo an inelastic scattering event as it escapes the solution. $E_{cut}$ correspondingly serves as a liquid-phase reference point for quantifying BEs. In the experiment, however, $E_{cut}$ is revealed by the large signal background of inelastically scattered electrons, whose signal intensity is *cut off* by the surface-barrier limit. Such measurements are routinely performed in solid-state systems but were only performed with aqueous solutions many years after the invention of the volatile-liquid-microjet technique in 1997[3] and the early development of the LJ-PES research field in approximately 2004[4]. Although the first measurement of $E_{cut}$ was reported as early as 2003,[5] the approach was only recently re-introduced[6, 7] and accurately applied.[1]

The reasons for this sluggish development were recently reviewed in detail by some of the authors.[1] So far, the LJ-PES community largely relied on known reference photoelectron peak BEs in the respective solvent signal to determine other liquid-phase BEs. One rather involved method to achieve this is to use gas-phase signals to determine liquid-phase BEs, with the former having well-known BEs and inevitably appearing in the spectrum together with the liquid-phase signals due to evaporation from the target. The main complication with this practice is that the surface charge of the liquid jet is difficult to quantify, and as a result, the energy calibration of a measured liquid-water photoelectron peak with respect to the corresponding and known gas-phase ionization energy is only approximate. The error depends on the degree of surface charge, which can vary widely from solution to solution, and hence on the magnitude of the electric field between the liquid jet and the grounded electron detector. Accordingly, liquid-phase (nearly) neat solvent peak BEs have been carefully pre-calibrated using the aforementioned methodology and subsequently used as liquid-phase energy references for aqueous solutions, assuming that the reference solvent BEs are invariant with solute concentration. Here, water's highest occupied molecular orbital (HOMO) $1b_1$ band BE has acted as the reference for the valence spectral region, with the O 1s BE being regularly used as a BE reference for core-level spectra. The simplicity of this approach resulted in it becoming a convenient and well-established, although flawed, BE calibration procedure for LJ-PES. A major associated consequence is that all PES studies from aqueous solutions to date, other than our own recent study,[1] did not and could not measure the solute-induced effects on the lowest $1b_1$ ionization energy of liquid water, or absolute-energy-scale changes to its electronic structure more generally. Hence, the systematic errors of previously reported solute ionization energies have the potential to be substantial, particularly when high bulk or local solute concentrations are implemented. Indeed, electrolytes are expected to induce significant electrostatic effects and disruptions of the hydrogen-bonding network in liquid water, particularly for highly concentrated solutions (see Ref. [8] and references therein), where the iodide anion has been reported to have an especially large influence on the extended hydrogen-bonding network.[9] In such solutions, the highly unsatisfactory situation of being unable to quantify any possible energy shifts of the water $1b_1$ orbital energy, and absolute-scale water electronic energetics in general, has been accentuated only recently[8] after decades of LJ-PES research. However, with the additional determination of $E_{cut}$ in LJ-PES experiments, such measurements now become possible and BEs of both solvent and solute can be determined absolutely, without assumption, and without relying on a gas-phase-referencing method.

An equally important and recent LJ-PES methodology development permits the accurate determination of surface properties of liquid solutions, such as work functions ($e\Phi_s$).[1] Thus far, the LJ-PES community has largely neglected the characterization of such surface properties, with just four exceptions.[1, 6, 7, 10] In fact, this field of research has largely been discussed within the domain of molecular physics. However, in order to explicitly account for the liquid surface and accurately determine liquid-phase BEs and surface potentials, condensed-matter concepts must be invoked, as further demonstrated here.

In the present study, we apply the new experimental tools discussed above to quantify the solute-induced evolution of water's valence electronic structure and the lowest ionization energy of the solute. This is exemplified via concentration-dependent NaI and tetrabutylammonium iodide (TBAI) aqueous solution LJ-PES measurements, spanning dilute to near-saturated bulk and supersaturated solution concentrations, respectively. The solubility limit of NaI in water is ~12.3 M at room temperature,[11] and at such high concentrations there are approximately only five water molecules per $Na^+/I^-$ pair, implying extensive solution-structure and composition changes, as well as ion pairing. An associated expectation is that such bulk-solution structure modifications would be reflected in the liquid water and iodide ($I^-$ 5p) valence PES spectra, as explored in an earlier work by some of the authors.[8] This previous study reported similar LJ-PES measurements to those reported here, also spanning concentrations between 0.5 M and 8.0 M but recorded with photon energies of 180 eV and 650 eV using a synchrotron radiation source. In the present study, a laboratory 40.814 eV (He II α) photon source is alternatively implemented, with no observable effect on the relative peak energetics extracted in the previous study. In fact, the previous PES spectra are almost replicas of those to be presented here, except for the relative spectral signal intensities arising from photon-energy-dependent photoionization cross-sections. Yet, the decisive difference is that we now also measure $E_{cut}$ from which, together with the accurately known photon energy, absolute solvent and solute binding energies can be accurately determined.[1] In the previous Pohl et al. study,[8] on the other hand, the PES spectra measured at different NaI concentrations were aligned at the positions of the water $1b_1$ peaks. This approach was justified by the fact that the entire photoelectron spectrum experienced an average uniform energetic shift as if a bias voltage had been applied to the sample. With that, any signature of electronic structure change could be quantified solely with respect to a fixed water $1b_1$ energy. Nevertheless, the Pohl et al. study did reveal a number of water-orbital-specific, relative energy changes that were interpreted with the help of high-level electronic-structure calculations. One of the conclusions from the theoretical data was that the liquid water $1b_1$ peak position essentially remains unaltered (*i.e.*, changes were very small) with increasing electrolyte concentration, which to some extent would justify the experimental energetic referencing procedure. This latter aspect partially explains the particular interest in water's absolute $1b_1$ BE, as well as the fact that this peak is generally well isolated in the photoelectron spectrum for both the liquid- and gaseous phase. More importantly, this energy is a determining factor for chemical reactivity with the solvent in aqueous solution.[12] Yet, regarding the molecular structure of liquid water more generally, a particularly sensitive fingerprint is the water $3a_1$ PES peak shape, which will also be addressed here. This is connected with a pair of orbital components that are primarily associated with intermolecular bonding and anti-bonding interactions between water molecules, and represent orbitals that are affected by explicit water–water and ion–water interactions. Shining new light on exactly these aspects is a major goal of the present work. A secondary aim is to provide more accurate absolute-energy-scale experimental water $1b_1$ and $3a_1$ as well as iodide $I^- 5p_{(aq)}$ BEs to enable a direct comparison between measured experimental data and the results of high-level electronic-structure theory and associated spectral simulations.

In the case of the TBAI surfactant – where bulk-solution concentrations are much lower but sufficient to achieve surface (super)saturation – we may expect that water $1b_1$ energies correlate with the formation and magnitude of a molecular surface dipole. There may also be correlations of surface-dipole effects with the $I^- 5p$ energy. Hence, the crucial difference between the NaI and TBAI systems is that the latter will allow primary and specific exploration and quantification of $e\Phi$, an explicit surface property, from an aqueous solution. In fact, the present $TBAI_{(aq)}$ study, performed using the new experimental capabilities, can be compared to one of the very early LJ-PES studies[4, 5] – also on TBAI – where experimental conditions did not permit the current questions to be addressed. Further interest in this particular surfactant system arises from its use as a highly efficient phase-transfer catalyst.[13]

Since our experiments aim at the characterization of the solution interface with concurrent sensitivity to the bulk of the solutions, a sufficiently large probing depth of our generally surface-sensitive method must be assured. At the photon energy of 40.814 eV applied here, the leading water valence photoelectrons have a ~30 eV KE, which is thought to correspond to a 1-2 nm electron inelastic mean free path (eIMFP) in neat water.[14-16] Corresponding values for the solutions are not accurately known but we assume that the experiment probes several layers into solution, exponentially attenuated with the IMFP (or more precisely, an effective attenuation length)[17] for a given electron KE; probing depth in 10 M NaI aqueous solution has been estimated to decrease by ~30% as compared to pure water at 65 eV KE.[18] The aforementioned length scale is well-matched to that over which bulk conditions pertain in aqueous salt solutions.[19] Indeed, the similarity of the relative $NaI_{(aq)}$ solution energetics reported here and previously at significantly higher photon energies of 180 eV and 650 eV,[8] corresponding to 1-4 nm probing depths,[14-16] indicates that the photon energy implemented in this study provides sufficient depth sensitivity to interrogate the interface and bulk solution behavior. One other crucial aspect is that the 40.814 eV photon energy is large enough to produce valence photoelectrons with energies larger than a threshold KE of approximately 10-15 eV below which quasi-elastic electron scattering in solution causes peak distortions, and binding energies can no longer be determined.[2]

## II. Experimental

All photoelectron experiments were performed with the EASI setup.[20] It comprises a state-of-the-art near-ambient-pressure capable hemispherical electron analyzer (HEA, HiPP-3, Scienta-Omicron) which detects electrons generated upon ionization of a 28-μm diameter liquid jet formed from a glass capillary at the exact photon energy of 40.814 ± 0.002 eV. This energy is provided by a VUV laboratory He-discharge light source (VUV5k, Scienta-Omicron), with the emission line being selected and pre-focused via a curved diffraction grating. The discharge lamp emits essentially unpolarized light which is only minimally polarized (<0.1%) by the monochromator system as it is delivered to the LJ. The photon-energy resolution was limited by the intrinsic width of the emission line, He II α, of 2 meV. After the monochromator, the light is further collimated via an exit capillary down to a focal spot size of approximately 300 x 300 μm² at the LJ sample. The light propagation axis spanned an angle of ~70° with respect to the photoelectron detection axis; LJ propagation and photoelectron detection axes were orthogonal to each other. The electron analyzer resolution was better than 40 meV at a pass energy of 20 eV. For all measurements, we used the so-called VUV lens mode. In this work, we were mainly interested in detecting the water $1b_1$ and I⁻ $5p$ photoelectron peaks and the low-energy tail, including $E_{cut}$. Measurement of the latter necessitates the application of a negative bias voltage at the jet, -25 V for all measurements reported here. This separates the cutoff energy of the solution from that of the electron detector. A beneficial side effect is that liquid-phase spectra can be obtained with nearly no gas-phase contributions.[1] Liquid jet biasing is accomplished by placing a metallic tube in between the high-pressure liquid PEEK lines that feed the glass capillary. This piece, which is thus in direct contact with the liquid approximately 55 cm upstream of the capillary, can either be electrically connected to the grounded HEA or to a highly stable Rohde & Schwarz HMP4030 power supply.

Liquid flow rates for all solutions other than 8 M NaI$_{(aq)}$ were set to 0.8 ml min⁻¹, which translates to an approximately ~20 ms⁻¹ jet velocity. For 8 M NaI$_{(aq)}$ we used 1.2 ml min⁻¹ (~30 ms⁻¹) to maintain better jet stability. The solution bath temperature, as regulated by a chiller unit, was typically 10°C for all solutions other than 8 M NaI$_{(aq)}$, where we used 15°C to avoid precipitation. Upon injection into vacuum, the LJ is formed with a laminar flow region extending over 2–5 mm, which is positioned ~800 μm away from the HEA entrance aperture, with an 800-μm entrance aperture diameter. The jet was ionized right in front of the HEA. At this short distance, electrons emitted from the liquid phase can reach the differentially pumped electron-detection chamber unperturbed at an increased transfer length of ~1 mm under typical experimental conditions. The average pressure in the interaction chamber during liquid-jet operation was approximately 7×10⁻⁵ mbar, accomplished with two turbo-molecular pumps (with a total pumping speed of ~2600 L/s for water) and three liquid-nitrogen cold traps (with a total pumping speed of ~35000 L/s for water). The pumping speed, $S$, per surface area (*i.e.*, in L/s/cm²) of the latter was estimated as $S = 3.64 \sqrt{T_{gas}/M}$,[21] where $T_{gas} \approx 273$ K is the temperature of the water vapor and $M = 18$ is the molar mass of water, which yields $S \approx 14.2$ L/s/cm². Experimental details, including collection of the liquid and emerging droplet spray, jet fine-positioning, relevant HEA features, and vacuum pumping system are described in Ref. [1]. Aqueous solutions were prepared by dissolving NaI or TBAI (both Sigma-Aldrich and of +99% purity) in highly demineralized water (conductivity ~0.2 μS cm⁻¹) and were degassed in an ultrasonic bath for ~5-10 minutes. The solution was delivered using a Shimadzu LC-20 AD HPLC pump that incorporates a four-channel valve for quick switching between different solutions. The equipped in-line degasser (Shimadzu DGU-20A5R), which is connected between the sample reservoir and the low-pressure side of the HPLC pump, was used as well during operation.

## III.   Results and Discussion

### III.1 Near-$E_{cut}$ and valence PES spectra from NaI aqueous solutions as a function of concentration

Fig. 1 presents PES spectra from NaI aqueous-solution microjets for several concentrations spanning 50 mM to 8.0 M; the lowest concentration of 50 mM is added to maintain sufficient conductivity for PE experiments but is otherwise considered indistinguishable from neat water.[4] Measurements were made from a 28 μm diameter liquid jet, biased at -25 V, and using a photon energy of hυ = 40.814 ± 0.002 eV. Fig. 1A presents the high-resolution LETs of the photoemission spectra with the characteristic low-energy cutoff, where we have applied the tangent method to plot the spectra on a common, bias-corrected KE scale (where $E_{cut} = 0$ eV; see the Introduction) and calibrated BE scale for the valence region;[1] signal intensities are normalized to yield the same cutoff slope. The BE scale is established via the relation $BE = h\upsilon - KE$. We note that this equation implicitly uses the spectral width, $\Delta E_w$, to determine the KE term, which we define as the energy distance from $E_{cut}$ to the PE feature of interest, *i.e.*, $\Delta E_w = KE_{measured} - E_{cut}$. If $E_{cut}$ is not aligned to zero beforehand, then rather $BE = h\upsilon - \Delta E_w$. Corresponding valence spectra are plotted in Fig. 1B, where the displayed spectral range covers the water $3a_1$, $1b_1$, and spin-orbit split iodide I$^-$ $5p_{3/2,1/2}$ doublet[8, 12, 22] signals occurring at KEs (bottom axis) of ~26-28 eV, ~29-30 eV, 31-34 eV and electron BEs (top axis) of ~13-15 eV, ~11-12 eV, 7-10 eV, respectively. Signal intensities are normalized to the $1b_1$ peak height for better visual comparability. As-measured spectra are shown in Fig. SI-1 in the Supporting Information; the maximal signal intensities were $\sim 0.4 \times 10^6 - 1.2 \times 10^6$ counts per s for the cutoff region and $\sim 0.4 \times 10^4 - 1.0 \times 10^4$ counts per s for the valence band, respectively.

The series of spectra shown in Fig. 1B is analogous to the respective data presented in Pohl et al.,[8] with the insignificant difference that at the lower $h\upsilon$ used in the present study, relative differences in ionization cross-sections yield somewhat larger water $1b_1$-to-I$^-$ $5p$ and $1b_1$-to-$3a_1$ signal intensity ratios. Another difference, which can be considered an improvement for a detailed analysis, but is otherwise irrelevant in the present context, is that the PES spectra in Fig. 1B contain no gas-phase water signal contributions. The reason is that the applied bias voltage between the liquid jet and detector orifice causes a potential which increases with distance, and only partially accelerates electrons liberated from the gaseous species some distance away from the liquid surface. As a result, the gas-phase signal is energetically smeared out and separated from the liquid-phase signal.[1] At most, the gas-phase contribution adds a broad background to the biased spectrum which, however, was negligible in our experiments. Following the spectral evolution, from the lowest to highest salt concentration, one observes an increase of the I$^-$ $5p$ signal intensity. However, the important finding is that the position of the $1b_1$ peak (and, on closer inspection, the I$^-$ $5p$ peak; see below) is not constant in energy, exhibiting a ~260 meV total shift towards lower KEs (higher BEs). Furthermore, a significant change of the water $3a_1$ peak shape is observed, arising from weakened intermolecular $3a_1$–$3a_1$ interaction upon addition of salt.[8] In the coming sections of this manuscript, we will present and discuss the absolute values of the various orbital binding energies for the two salt solutions and the respective concentrations. It is convenient (and consistent with a previous notation)[1] to interchangeably refer to vertical ionization energies, VIEs, which are a measure of the propensity to detach an electron under equilibrium conditions and are equivalent to the (vertical) binding energies. In both cases, the measured energy is related to the position of the maximum of the respective photoelectron peak. Thus, the $1b_1$ BE from neat liquid water is the same quantity as VIE$_{1b1,water}$, the water $1b_1$ BE from solution corresponds to VIE$_{1b1,sol}$, and the analogues for the water $3a_1$ and iodide I$^-$ $5p$ BEs are VIE$_{3a1,sol}$ and VIE$_{I5p,sol}$, respectively, with the subscript 'sol' either referring to NaI or TBAI aqueous solutions.

To extract the quantitative evolution of individual spectral features, concerning both peak position and area, we employed a fit with 4-6 peaks (4 peaks for the neat water spectrum where the I$^-$ $5p$ signal is absent). Gaussians were used for all peaks other than the $1b_1$ peak. We find that for spectra measured with sufficiently high resolution, as employed here, the simplified assumption of a 'Gaussian' $1b_1$ peak shape is insufficient to describe the asymmetric peak shape correctly. The asymmetry arises from vibrational structure which is not resolved in the liquid-phase spectra due to inhomogeneous (configurational) broadening;[23] see Fig. 3 for an exemplary water gas-phase valence photoelectron spectrum. We thus opt to describe the $1b_1$ peak by an exponentially modified Gaussian shape,[24] where the asymmetry $\tau$ is fixed to a value of -0.3 eV; asymmetry values of -0.2 to -0.3 have been found to describe the spectral envelope of the gaseous $1b_1$ peak well. The $3a_1$ split feature is constrained to yield the same height and width for both Gaussians.[4, 8] Furthermore, the I$^-$ $5p_{3/2}$ and $5p_{1/2}$ double peaks were constrained to yield the expected 1:2 peak area ratio. Exemplary fits, for all peaks, are plotted in Fig. SI-3, and the fit results are summarized in Table 1. The analogous analysis has been performed for the TBAI aqueous solutions (discussed later), and the results are summarized in Table 2.

Before quantifying and interpreting the observed energy shifts in Fig. 1B, one important conclusion that can already be drawn at this point is that all effects primarily reflect bulk-solution properties. This is inferred from the water $1b_1$ and I⁻ 5p signal intensities, specifically the areas from the peak fitting, as a function of concentration (bottom axis), as shown in Fig. 2A. The as-measured $1b_1$ signal intensity (black open triangles) is seen to monotonically decrease over the entire concentration range (also compare Fig. SI-1), while the relative, *i.e.*, $1b_1$-peak-area normalized, I⁻ 5p signal (red full triangles) monotonically increases. Such a quantitative balance results from the decreasing number of water molecules and the increasing number of ions in a given probing volume as the solute concentration is increased, which is possibly accompanied by increased electron scattering that further diminishes the water signal. That said, it is well established that heavier halide anions preferably accumulate at the liquid interface, with iodide being pushed out of the water network due to its large size and polarizability, resulting in a particularly high halide ion surface activity.[18, 19, 25, 26] The cation is correspondingly pulled towards the interface and a surface concentration enhancement is established for the two ionic species, with characteristic peaked, but slightly offset, density profiles.[25] We attempt to quantify the observed I⁻ peak intensity increase using the well-known BET (Brunauer, Emmett, and Teller) isotherm,[27] which was developed for multi-layer gas adsorption but has been shown to be equally applicable to describe water activity in concentrated electrolyte solutions by viewing hydration as an adsorption of multiple water shells around the electrolyte.[28] Here, we repurpose the equation to describe the buildup of ion concentration as an irregular 'multi-layer adsorption' at the water interface:

$$\text{Int}_{I5p} = \text{Int}_{sat} \frac{cX}{(1+X)[1+(c+1)X]} \tag{1}$$

with $X = [c]/[c]_{sat}$ being the fractional bulk-solute concentration to saturation concentration (which is about 12.3 M for $NaI_{(aq)}$ at room temperature)[11] ratio and the BET parameter $c = \exp(\Delta\epsilon_{ads}/RT)$ relating to the energetics of adsorption $\Delta\epsilon_{ads}$ in relation to the product of the gas constant R and temperature T, *i.e.*, the thermal energy. Here we assume $RT$ = 24 meV (0.562 kcal/mol, at 10°C). $\Delta\epsilon_{ads} = E_1 - E_L$ is composed of the heat of adsorption at the interface $E_1$ and the heat of liquefaction/vaporization $E_L$, representing the strength of interaction of the adsorbing species with the interface and with itself, respectively. $\text{Int}_{I5p}$ and $\text{Int}_{sat}$ is the observed and the maximal expected intensity of the I⁻ 5p peaks, respectively. In our context, the 'adsorption' (interface enrichment) happens at the liquid-vacuum interface and is driven by the increase in bulk concentration. Even though we primarily concern ourselves with the I⁻ peak intensity here, as the Na⁺ peaks are severely perturbed or not observable at the photon energy implemented here, an analogous behavior is expected for the cation. Na⁺ is pulled towards the interface by the attraction of the anion, *i.e.*, both anion and cation intensities increase in unison[22] and the anion peak intensity in our analysis is representative of the behavior of both species. A fit of the concentration-dependent iodide-5p-to-water-$1b_1$ peak-area ratio data shown in Fig. 2A to Eq. 1 yields an excellent match to the data, with a value of $c \sim 4.2 \pm 0.6$ being extracted, corresponding to $\Delta\epsilon_{ads} = 0.8 \pm 0.08$ kcal/mol. This value is comparable to, *e.g.*, adsorption of cold (90 K) nitrogen on various surfaces such as silica gel,[27] and hints at a moderate-to-low, unfavorable buildup of ion density at the interface. A more detailed analysis of this behavior is beyond the scope of this work, however. More importantly for the following discussion, we argue that the observed surface enrichment does not lead to a significant change to liquid water's nascent surface dipole and/or a buildup of an appreciable surface dipole perpendicular to the surface, *i.e.*, the additional solute charges are largely compensated in the perpendicular direction. In particular, we emphasize that an interface enrichment is necessarily followed by ion depletion in the subsurface region so as to maintain thermodynamic equilibrium, and the net effect is still a lower ion concentration in the overall interfacial region.[25] Thus, any differential segregation, implying the formation of an electric double layer (separating the anions and cations by approximately 3 Å), is counter-balanced by the subsurface, and the net effect is that the majority of photoelectrons (born in deeper layers) only experience a minor deceleration field. This will be detailed below when we discuss the analogous, but very different, results from $TBAI_{(aq)}$.

Fig. 2B presents the quantitative evolution of $VIE_{1b1,NaI}$ (blue open circles; left axis). At the lowest salt concentration $VIE_{1b1,NaI} = VIE_{1b1,water} = 11.33$ eV,[1] with this value increasing to 11.6 eV at the highest concentrations. The associated error bars are small, and are included in the figure; the highly precise values presented here are a result of using consistent, high-resolution settings throughout the whole measurement series. The VIE increases analogous to the trend of interfacial ion concentration, *i.e.*, the I$^-$ peak signal, with an essentially linear increase until approximately 4 M concentration, followed by a somewhat steeper rise towards higher concentrations. This trend, and the 1:1 correspondence to the I$^-$ peak-signal increase, is confirmed when comparing the BET curve from panel A with the change in $VIE_{1b1,NaI}$ in panel B; we reproduced this curve in blue which was scaled / offset as a visual guide. Again, an excellent match is observed. One may speculate that at higher bulk concentration, *i.e.*, where the interfacial ion concentration rises rapidly, a major solution structure-change occurs, which would seem plausible since the associated water-to-ion ratio is approximately 7:1. With a reported water hydration-shell number of 8 for I$^-$,[29, 30] and 4.5-6.0 for Na$^+$,[29] this 4 M concentration coincides with an increasing probability of solvent-shared hydration configurations. Indeed, theoretical calculations reveal an increasing number of solvent-shared ion pairs and contact-ion pairs – see, *e.g.*, Ref. [8, 31] –, and noticeably the total fraction of ion-pair structures increases significantly when passing from 3 M to 8 M solution.[8] At 3 M concentration, for instance, the coordination numbers of the ions around water are 0.450 for an iodide anion and 0.329 for a sodium cation. Also, the water structure is slightly altered in the 3 M solution,[8, 31] assuming less tetrahedral character compared to bulk water. More dramatic effects occur for the 8 M solution, judged from the distance of the closest water molecules, quantified by the O–O radial distribution functions. The observed +260 meV energy shift (Fig. 2B) can be compared with the +200-meV calculated shift in Fig. 9 of Ref. [8]. To be more specific, the calculations find a <100 meV energetic shift relative to $VIE_{1b1,water}$ when going from zero to 3 M concentration, and the effect increases to ~200 meV, corresponding to $VIE_{1b1,NaI} = 11.53$ eV, when going to 8 M. Arguably, this observation appears to coincide with the steeper energetic changes (Fig. 2B). We thus find that the observed energy shift is almost fully explained by electronic structure changes, and, considering the very small discrepancy to theory (~0.07 eV at 8 M), a change in the solution's work function is, if occurring at all, very small. The remaining discrepancy may well be explained by a small solution surface-dipole change at very high concentrations, originating from a charge imbalance perpendicular to the interface and/or reorientation of water molecules driven by the present surface charge. In the former case, the dipole between I$^-$ directly at the surface and Na$^+$ in the immediate sub-layer leads to a somewhat higher $e\Phi$, which translates to a small additional increase in $VIE_{1b1}$ at very high concentrations. It is interesting to note that the remaining discrepancy of ~70 meV agrees well in absolute value *and* direction with the change in surface potential of about ~40-50 mV when going from neat water to highly concentrated $NaI_{(aq)}$ as reported by Nguyen et al.[32] However, considering the assumptions made and error intervals involved, we are unable to draw any definitive conclusions here. Regarding the overall slight changes of $VIE_{1b1}$, we conclude that fixing the $1b_1$ energy, as done in the Pohl work, with the aim of determining solute BE turns out to work rather well in the case of $VIE_{I5p,NaI}$. It does not mean, however, that fixing the $1b_1$ energy is a generally valid approach; $TBAI_{(aq)}$ in fact will be shown to exhibit a very different behavior. The reason for the (unexpected at the time before publishing Ref. [8]) small $1b_1$ energy change in the case of $NaI_{(aq)}$, despite the transitioning from essentially hydrogen-bonded neat liquid water to crystalline-like liquid phase, has been attributed to an isolation and stabilization of the non-bonding $1b_1$ electron by the charge-dense sodium cation. This is accompanied by the destabilization of the water $1b_2$ electron by the iodide anion, which is not considered in the present study. Pohl et al. have also discussed the possible effect of concentration-dependent variations of the dielectric constant on $VIE_{I5p,NaI}$, but establishing such a relationship requires additional experimental studies.

Associated $VIE_{I5p,NaI}$ are plotted in Fig. 2B (green color; shown for I$^-$ $5p_{3/2}$). The $VIE_{I5p,NaI}$ energy shift is also linear, of almost the same magnitude as for $VIE_{1b1,NaI}$, and exhibits a similar small departure from linearity at the same 4 M concentration, but this time the energies increase at a slightly slower rate. We emphasize that any possible change in the solution streaming potential with solute concentration is irrelevant here, the additional potential would simply add to that associated with the bias voltage and equally offset the spectral cutoff and valence ionization features used to calculate the VIEs, see the beginning of section III.1 and Ref. [1] for details. Hence, since there is no obvious experimental reason that could cause the observed opposing trends of the two independently measured quantities, we once more corroborate the occurrence of structure changes that are reflected in the PES spectra of both the water solvent and the iodide anion. It is noted that the $VIE_{I5p,NaI}$ values in the present study (Fig. 2B) are somewhat larger than found in Ref. [8], which simply arises from the fact that $VIE_{1b1,NaI} \neq VIE_{1b1,water}$. A quantitative understanding of the $VIE_{I5p,NaI}$ shifts must await theoretical calculations and, at this point, we conclude with a previous statement that the shifts might be caused by the electrolyte-induced hydrogen-bonding network disruption and associated changes in charge donation by the polarizable $I^-$ anions to the water anti-bonding, $\sigma^*(O-H)$, orbitals as the electrolyte concentration is increased.[9]

Aforementioned solute-induced effects on the water $3a_1$ peak shape will be only briefly addressed here because the findings are exactly the same as reported in Ref. [8]. Furthermore, the analysis largely concerns the quantification of the energetic split of the two $3a_1$ components, and absolute energetics provides marginal new information on this particular aspect. Nevertheless, it is useful to present the data here for a direct comparison of the analogous measurements from $TBAI_{(aq)}$ where the hydrogen-bonding network and its changes would be expected to play a minor role. For that, we recall the origin of the water $3a_1$ flat-top spectral profile (see also Fig. SI-3C), which is typical for neat liquid water, and what causes its narrowing and the observation of a broad peak maximum when the NaI concentration is increased. This can be readily seen in Fig. 1B. The flat-top shape in the case of neat water results from the contribution of two $3a_1$ orbitals, each of which can be represented by a Gaussian of the same width and height, at BEs of $13.09 \pm 0.05$ eV and $14.53 \pm 0.05$ eV for neat water, primarily associated with intermolecular bonding and anti-bonding interactions between water molecules. The lower-BE-energy peak is referred to as the $3a_1$ L band and the other contribution as the $3a_1$ H band. The (nearly) neat water $3a_1$ peak splitting reduces by $450 \pm 90$ meV for the highest NaI concentration, as can be seen in Fig. 2C, with the decreasing energy splitting causing the observed change of peak shape, in excellent agreement with Ref. [8]. Fig. 2C suggests a linear decrease of the peak splitting and, as in Fig. 2B, there might be an indication of departure from linear behavior near a concentration of 4 M. Such $3a_1$ H - $3a_1$ L energy narrowing, upon addition of salt, has been attributed to weakened $3a_1$–$3a_1$ intermolecular electronic interactions, modulated through the replacement of water units by ions.[8]

Related to the decrease of the quantitative water–water hydrogen-bonding interactions for sufficiently high NaI concentrations already addressed above, we present another spectral analysis, based on two experimental observables, which descriptively map the evolution from the water gas-phase spectrum into the 8 M NaI solution spectrum. We start with the well-studied gas-phase water spectrum, shown in Fig. 3 (grey-dotted curve), here presented on the KE scale, as measured in the experiment. Note that such a spectrum can be readily measured from the water gas-phase molecules near the liquid jet, where their density is largest. For that, the liquid jet is slightly moved downwards so that the VUV light barely intersects with the liquid. Vibrational resolution, as achieved here, is however only possible if the liquid jet surface is not charged which corresponds to electron detection under field-free conditions; this has been discussed in great detail in Ref. [1]. The respective neat water liquid-phase spectrum, black-dashed curve, has been simulated by convolution of the gas-phase spectrum with a Gaussian of FWHM = 1.45 eV, in accordance with the liquid $1b_1$ peak width reported in Ref. [4], and shifted by 1.02 eV to higher electron KEs (lower eBEs), which corresponds to the gas–liquid shift of 1.28 eV (12.62 eV- 11.34 eV)[1] for neat liquid water, and corrected by the 0.26-eV shift after adding 8 M NaI. Furthermore, a simple Shirley-type background[33] has been added to include the effect of inelastic scattering for better comparability. Our simple modification of the water gas-phase spectrum solely accounts for the unspecific structural inhomogeneity, i.e., peak broadening due to the statistical distribution associated with different configurations, and polarization screening inside the liquid environment, associated with an empirical change of the dielectric function when going from water to highly concentrated NaI. The result is found to be in an excellent agreement with the 8 M $NaI_{(aq)}$ spectrum. Having fully neglected any hydrogen-bonding-specific effects in our simple modeling approach, Fig. 3 directly shows that water hydrogen bonding in the 8 M solution is absent or at least vastly reduced. Furthermore, our data provide the necessary energetic information against which theoretical modeling of concentration-dependent dielectric functions can be gauged, as was also alluded to in Ref. [8]. Another noteworthy implication of our comparison in Fig. 3 is that the same peak widths, which are characteristic of inhomogeneous structural broadening in neat liquid water, can be used to model the 8 M solution spectrum. It seems that the energetic distribution in the fluxional hydrogen-bonding network is balanced by inter-ionic interactions in the more viscous environment.

We conclude the section on NaI solutions by inspecting the LET shape over a wider range, up to 8 eV above $E_{cut}$; see Fig. 4. At low salt concentrations, this distribution exhibits a rather broad, approximately 2-eV wide, structureless peak with a maximum near ~0.8 eV from $E_{cut}$. This is the typical shape observed for neat liquid water.[2] Upon increasing the concentration, this peak narrows and its maximum shifts closer to $E_{cut}$, and this is accompanied by an edge evolving near 5 eV KE; it seems that the two effects are quantitatively balanced. This spectral evolution is, however, unrelated to the electronic structure aspects that we are interested in, but is of interest for a different reason: It relates to a comment earlier in this paper on the ability to extract accurate binding energies if the respective photoelectron peak is at a KE less than 10 eV. Then, strong quasi-elastic scattering leads to a build-up of a broad signal background at the position of the associated photoelectron peak.[2] Qualitatively, this is exactly what we observe in Fig. 4, however, with the new aspect that electron scattering is now probed in highly concentrated aqueous solutions, where the probability of electron scattering with atomic ions is large, and dominating at the very large concentrations. Theoretical modeling of LET shapes, containing information on the active scattering mechanisms and their probability, from both neat liquid water and aqueous solutions is an ongoing challenge.[2] The specific photoelectron peak that occurs near the origin of the photoemission spectrum is associated with the $Na^+$ $2p_{(aq)}$ ionization channel, with ~35 eV BE.[8, 22, N1] This poses an intriguing example of the strongly enhanced quasi-elastic scattering in the <10-13 eV region, in addition to the cases presented in Ref. [2]. The $Na^+$ 2p is a particularly strong signal, easily dominating the spectrum at high concentrations,[22] which enables us to directly observe the deterioration in shape of a mostly (initially) Gaussian-shaped PE feature, in addition to the inevitable reduction in signal intensity. So far, the presented examples in Ref. [2] had a rather small intensity to begin with, which made a close inspection of the peak shapes after suffering strong quasi-elastic scattering difficult. In Fig. 4, one can clearly observe that the initial peak with a FWHM ≈ 1.23 eV is smeared out to a broad plateau spanning more than 4 eV. It can be expected that all PE features will be distorted in a similar way, reinforcing our conclusion in Ref. [2] that PE features in liquid water and aqueous solutions cannot be reasonably extracted below an electron KE of 10-13 eV.

**III.2 Near-$E_{cut}$ and valence PES spectra from TBAI aqueous solutions as a function of concentration**

Figures 5A and 5B present LETs and associated valence PES spectra from a microjet of a TBAI aqueous solution for several concentrations, 0 to 40 mM. Experimental conditions were the same as stated above when presenting analogous results from $NaI_{(aq)}$; we also applied the same bias voltage of -25V and display the same spectral ranges. LETs (Fig. 5A) are again presented as normalized to yield the same cutoff slope and aligned to $E_{cut}$ = 0 eV. The zero position of the energy scale, $E_{cut}$, then determines the KE position of the individual valence spectra (Fig. 5B). Qualitatively, the spectra are rather similar to the ones from $NaI_{(aq)}$ (Fig. 1B), exhibiting the water $3a_1$ and $1b_1$, and the iodide 5p photoelectron features. A major difference, most directly reflecting the hydrophobic interactions between water molecules and the $TBA^+$ alkyl chains, is the much larger water-to-iodide signal-intensity ratio for a given concentration, corresponding to an effective segregation factor of approximately 70.[34] An even larger factor of 300 has been reported in an ionization threshold study by Watanabe et al.,[35] which, however, is most likely an artefact of the employed measurement and analysis method. We find that the $VIE_{I5p}$ is much lower for $TBAI_{(aq)}$ than that for $NaI_{(aq)}$, which will give a proportionally larger photoelectron yield for $TBAI_{(aq)}$ in the 7.0-7.8 eV photon-energy range used by Watanabe et al. The shift in threshold energy with increasing TBAI concentration was observed in their study, but apparently not correlated to an increased ionization probability for this species, which may well have led to an overestimation of the segregation factor.

The most important differences between the TBAI$_{(aq)}$ and NaI$_{(aq)}$ solution energetics are (i) the considerably larger spectral energy shifts of the former, which are in fact rigid shifts of the spectrum as a whole, that trend in the opposite direction to the concentration-dependent shifts observed for NaI$_{(aq)}$, towards larger KEs / lower VIEs, and (ii) the absence of a pronounced change of the 3a$_1$ peak shape. Our observations are quantified in Fig. 6, based on the peak-fitting analysis described above. Before detailing the energetics, we consider the evolution of the iodide signal intensity in Fig. 6A (red symbols), which reveals adsorption characteristics of a strong surfactant. Unlike in the case of NaI$_{(aq)}$ (Fig. 2A), the iodide signal intensity rises rather linearly up to approximately 20 mM TBAI concentration, and then turns over into another seemingly linearly growing regime with reduced growth rate. We will invoke a Langmuir isotherm adsorption model to describe the data below. The results of Fig. 6A are in full agreement with earlier reports.[5, 34] The initial near-linear iodide signal increase is attributed to the regime of sub-monolayer coverage, with the single segregation monolayer being completed near ~20 mM concentration. Subsequent shallower signal evolution arguably corresponds to the filling of remaining cavities within the surface layer, and likely some slight increase of bulk-ion concentration.[34] Qualitatively, this behavior is further reflected in the accompanying water-signal attenuation shown in Fig. 6A (black symbols), which is also in excellent agreement with the early studies[5, 34] and results from the successive replacement of interfacial water molecules by solute ions.

An arguably more accurate description can be garnered in terms of a Langmuir adsorption isotherm model, which allows the surface-adsorption behavior of the solute ions to be analyzed and the extraction of the Gibbs free energy of adsorption, $\Delta G_{\text{ads}}$. Here we used the Langmuir adsorption model adapted to aqueous electrolyte solutions, which has previously been successfully applied to surface-active species in solution:[36-39]

$$\text{Int}_{\text{I5p}} = \text{Int}_{\text{sat}} \frac{K c_{TBAI}}{K c_{TBAI} + c_W} \approx \text{Int}_{\text{sat}} \frac{c_{TBAI}}{c_{TBAI} + 55.5\text{M} \exp(\Delta G_{\text{ads}}/RT)} \quad (2)$$

Here, $K$ is the equilibrium constant for surface adsorption, $c_{TBAI}$ and $c_W$ are the bulk solute and water concentration, respectively, and $RT$ = 24 meV (0.562 kcal/mol, at 10°C) is again the product of the gas constant and temperature. Indeed, a good fit to the data is obtained with Eq. 2, shown as a solid red line in Fig. 6A, with the fit parameter $\Delta G_{\text{ads}}$ = -0.19 ± 0.01 eV/molecule (-4.4 ± 0.2 kcal/mol). Note that the surface-adsorbing species is TBA$^+$ in this case, with I$^-$ in the sub-layer drawn to the surface by the TBA$^+$ cations. The valence TBA$^+$ signal arises at BEs greater 10.5 eV, as will be further discussed below. A Langmuir fit to this TBA$^+$ signal (Fig. 6C) yields a similar $\Delta G_{\text{ads}}$ = -0.17 ± 0.02 eV/molecule (-4.0 ± 0.4 kcal/mol), which shows the expected simultaneous surface enrichment of both ion species; the larger error reflects the fact that the TBA$^+$ signal was extracted from difference spectra with greater associated uncertainties in relative scale. Both values are in good agreement with previous reports on TBAI$_{(aq)}$.[40] Notably, the value for TBAI$_{(aq)}$ is smaller than the $\Delta G_{\text{ads}}$ of -0.26 ± 0.01 eV/molecule (-6.1 ± 0.2 kcal/mol) and -0.27 ± 0.01 eV/molecule (-6.3 ± 0.2 kcal/mol) observed for similar bulk concentrations of NaI$_{(aq)}$ and KI$_{(aq)}$, respectively (0-70 mM, as compared to our TBAI range of 0-40 mM), which was extracted from measurements of the I$^-$ charge-transfer-to-solvent (CTTS) transition *via* UV light from second-harmonic generation.[39] In the case of these simple, relatively low-concentration salts, the iodide ion is instead preferentially pushed to the surface, which leads to significant initial surface enrichment.

We next discuss the quantitative evolution of VIE$_{1b1,\text{TBAI}}$, VIE$_{3a1,\text{TBAI}}$, and VIE$_{I5p,\text{TBAI}}$, summarized in Fig. 6B and Table 2. It is seen that VIE$_{1b1,\text{TBAI}}$ = VIE$_{1b1,\text{water}}$ = 11.33 eV at zero TBAI concentration, with VIE$_{1b1,\text{TBAI}}$ decreasing to 10.60 eV at the highest concentration, 40 mM, which is a much larger energy shift and in the opposite direction than that of the NaI$_{(aq)}$ solute data shown in Fig. 1B. Furthermore, changes in VIE$_{1b1,\text{TBAI}}$ approach saturation with high concentration, as opposed to the $VIE_{1b1,NaI}$ data trend. In fact, the data indicate two different regimes, one below 20 mM and the other above that concentration, seemingly correlating with the adsorption curve of Fig. 6A, as will be detailed below. Analysis of $VIE_{3a1,TBAI}$ and $VIE_{I5p,TBAI}$ changes reveal the same energy shifts (within the error bars), presented in Figure 6B and Table 2, implying that the spectra rigidly shift as a whole, with no indication of differential behavior. These findings disagree with the earlier conclusion[34] that the water (as well as iodide) BEs do not change upon addition of salt but are qualitatively consistent with the interpretation in Ref. [5]. This discrepancy is connected to the problems with the gas-phase energy referencing used at the time, which insufficiently characterized surface charging, and a flawed measurement of $E_{cut}$ (where a bias voltage was used here to energetically separate $E_{cut}$ from the analyzer cutoff, unlike in Ref. [5]). In this context, we mention another PES study from 0.04 m (molal) $TBAI_{(aq)}$ microjets, reporting the VIE of the $I^-$ 5p from 0.04 m $TBAI_{(aq)}$ solution using gas-phase energy referencing.[41] The authors found somewhat higher values of $VIE_{I5p3/2} = 7.6$ eV and $VIE_{I5p1/2} = 8.4$ eV (no confidence interval was given) as compared to our results of $7.03 \pm 0.08$ eV and $8.0 \pm 0.1$ eV, respectively. The likely reason is a systematic error due to unknown and uncompensated extrinsic potentials from surface charging or the streaming potential as explained above and in Ref. [1].

Regarding possible effects on the water $3a_1$ peak shape, no narrowing of the $3a_1$ L – $3a_1$ H energy splitting is observed as opposed to $NaI_{(aq)}$. This may either imply that the electronic structure of the interfacial water molecules does not change (which is unlikely), or the effect is not detected over the probing depth of the experiment. Although the eIMFP is expected to be rather small (see the Experimental section), the largest fraction of the detected water signal apparently still comes from molecules with undisturbed electronic structure. What further complicates the analysis of the $3a_1$ peak shape (see Fig. 6C) is that this peak overlaps with a valence peak from $TBA^+$ (see Fig. 7), which is the reason for the observed overall signal intensity increase in the water $3a_1$ spectral region.[34] This prohibits accurate isolation of a potential small $3a_1$ peak narrowing. Hence, with the available experimental information, it remains unresolved whether TBAI has an effect on the water $3a_1$ orbital. The $TBA^+$ signal underlying the $3a_1$ peak is not considered in our fit, and we opted to constrain the $3a_1$ peak split to 1.4 eV (the value for neat water) in all fits to the $TBAI_{(aq)}$ spectra and instead report the peak width of the $1b_1$ peak as a function of concentration in Fig. 6C (full squares; left axis), assuming this to be exemplary for the overall broadening observed for all water features in the PE spectrum (also compare to Fig. SI-4 in the SI). We still attempted to isolate the $TBA^+$ signal contribution by taking the difference of the spectrum for each TBAI concentration with the spectrum of neat water. The resulting difference spectra are shown in Fig. 7, and the $1b_1$-to-$TBA^+$ peak-area ratio is shown in Fig. 6C as well (open squares, right axis). The $TBA^+$ signal has a rather large contribution to the valence spectrum (almost the same area as the $1b_1$ peak at a concentration of 35 mM) and increases in a similar way to the $I^-$ 5p signal (compare Figs. 6A and 6C).

Before discussing the origin of the observed changes in VIE in detail, we briefly comment on the overall LET shape, for which wide-spectral-range measurements are shown in Fig. 8. In contrast to the $NaI_{(aq)}$ results, no pronounced changes in LET shape are observed, with no solute PE features being expected in this energy region for the $TBAI_{(aq)}$ solution, which is apparent when comparing spectra measured at higher photon energies (see, *e.g.*, Fig. 1 in Ref. [34]). Upon close inspection, we find only a slight LET signal increase around an electron KE of ~1-4 eV. We speculate that this increase correlates with the scattered electron signal contribution from $TBA^+$ at a (bias-compensated) electron KE of ~25-30 eV (Fig. 7), where the most probable inelastic electron scattering occurs towards 20-25 eV lower electron KEs in water[42, 43], *i.e.*, into the 0-5 eV region of the spectrum.

What then is the reason for the large negative BE shifts, and what causes their apparent correlation with TBAI surface coverage? As we have seen, in the case of $NaI_{(aq)}$ the relatively small changes of $\Delta E$ likely primarily arise from electronic structure changes of the bulk solution, while interfacial molecular dipoles play a smaller role. Our earlier discussion of the $NaI_{(aq)}$ case highlighted charge neutrality preservation at the solution interface and the ion-density increase at the surface being (over-)compensated by a depleted sub-surface region, which results in an overall lower ion concentration in the interfacial region (see chapter III.1). In effect, photoelectron KEs are only minimally affected when traversing such an interface region. However, for surface-active TBAI$_{(aq)}$, the situation is very different. The high concentration of interfacial solute molecular dipoles will lead to work-function changes which are revealed as rigid spectral shifts, provided there is a considerable net dipole component perpendicular to the solution surface. Yet, the previous, aforementioned study by Watanabe et al.,[35] which determined concentration-dependent threshold ionization energies of I$^-$ 5p from TBAI$_{(aq)}$, solely attributed the observed energy shifts, of almost the same magnitude and sign as shown in Fig. 6B, to hydration changes and the decrease of iodide hydration number. Specifically, the authors found a rather complex multi-step variation of the threshold energy which was suggested to reflect the concentration-dependent stepwise dehydration of iodide (and stabilization by TBA$^+$) from a hydration number of six to four, three, and then two. Work-function effects were not considered, but are suggested here. We argue that they make the major contributions to the observed BE changes, as we explain in the next paragraphs. Note that concentration-dependent electronic structure changes of interfacial water, VIE$_{1b1,TBAI}$ and VIE$_{3a1,TBAI}$, have not been quantitatively discussed as of yet; respective computations are not available.

An experimental indication of significant e$\Phi$ effects is revealed from the inferred invariance of the water 3a$_1$ peak shape (point (ii) above) as well as a slight overall broadening of all spectral features, exemplarily shown for the 1b$_1$ peak width in Fig. 6C, as discussed above. With the aforementioned expected experimental probing depth and reduced fraction of interfacial water, with relatively small concentration, essentially remaining undetected on the large signal background from undisturbed bulk water, the large observed spectral shifts are deemed highly unlikely to arise from interfacial electronic structure changes. Indeed, recent 25 mM TBAI$_{(aq)}$ solution LJ-PES measurements have extracted a e$\Phi_{TBAI,25\,mM}$ value of 4.25 ± 0.09 eV and demonstrated a solute-induced e$\Phi$ reduction of 0.48 ± 0.13 eV with respect to nearly neat water.[1] We thus discuss how e$\Phi$ changes would play out, regarding both the magnitude of the energy shifts and their sign. Qualitatively, a decrease of e$\Phi$ by a negative surface dipole, $\varphi_{dipole}$, is associated with a dipole layer with negative charge pointing into the solution and positive charge residing at the top surface. This corresponds to the commonly assumed structure of the TBA$^+$I$^-$ segregation layer.[44, 45] An emitted electron is hence accelerated within this interfacial dipole field, acquiring a larger kinetic energy, consistent with the experiment (Fig. 6B). The effect scales with concentration, with the observed initial near-linear decrease of both VIE$_{1b1,TBAI}$ and VIE$_{I5p,TBAI}$ and increase of the respective KE suggesting that the dipole orientation varies insignificantly until the monolayer is completed. A slightly smaller energy shift of VIE$_{I5p,TBAI}$ is, however, barely quantifiable given the experimental error but would indeed be expected since the TBA$^+$, with its associated iodide counter ion, resides at the very top of the surface and should, hence, be less affected by an interfacial dipole layer. Smaller energy changes that occur at yet higher concentrations, corresponding to denser packing of the solute monolayer, can be associated with increasing dipole-dipole interactions. One might expect considerable variation of the relative position of iodide and TBA$^+$, as well as cation re-orientation, in an increasingly sterically hindered dense monolayer packing, but this is not supported by the experiment. With a maximum TBA$^+$ surface coverage of approximately $n = 1.3 \times 10^{14}$ cm$^{-2}$ (arguably corresponding to the completed monolayer near 20 mM concentration),[44] and $\Delta\varphi_{dipole,TBAI} = 0.7$ eV (from Fig. 6B), we can estimate an effective dipole moment of TBAI, using the Helmholtz equation $\Delta\varphi_{dipole} = \frac{enm}{\epsilon_r\epsilon_0}$,[46] where $e$ is the elementary charge, $m$ is the dipole moment, $\epsilon_0$ and $\epsilon_r$ are the vacuum and relative permittivity, respectively. This value can then be compared with the actual dipole moment of TBAI ($m = 13$ D, see Ref.[47]) to infer the average orientation of the dipole moment relative to the solution surface. With the values assumed here, and using $\epsilon_r = 1$ for the liquid-vacuum interface, we get $m = 1.43$ D, which is an order of magnitude lower than the actual dipole moment. The result strongly hints at a molecular arrangement largely in-plane of the solution surface with only a small component in the perpendicular direction, and the charge may be partially screened by the interaction with water. Such a behavior was also observed in MD simulations of 16 TBAI ion pairs in a water slab, where the orientation profiles for the butyl chains spiked at two angles, both of which are primarily in the interfacial plane.[34] It was further found that the water-induced and TBA$^+$-induced dipoles pointed in opposite directions, resulting in partial compensation. This is to be expected, as it is unlikely that the fluctuating solution interface would support an ordered, perpendicular arrangement of the TBA$^+$I$^-$ dipole, and the system rather is driven towards charge neutrality as far as possible.

A more assertive, although elusive approach to directly measure the concentration-dependence of the eΦs would be to experimentally determine the changes of $E_{cut}$, as often practiced in solid-state PE spectroscopy, *e.g.*, when assessing eΦ changes of overlayers atop metallic substrates.[48] This would require the simultaneous measurement of the system Fermi energy and the solution spectra, including the LETs under biased conditions, which is, however, elusive for the following reasons.[1] Water, a large-band-gap semiconductor,[49-51] does not exhibit a measurable Fermi edge itself (the electron density at the Fermi level / electrochemical potential is zero). Thus, the Fermi edge spectrum of an external (metallic) reference electrode in equilibrated electrical contact with the solution has to be measured separately. The problem lies in relating this external reference spectrum to the spectrum of the solution. This would require correct assessment of the different bias voltages actually applied to the reference electrode and the liquid (the liquid has additional internal resistances) and of the additional extrinsic potentials such as the streaming potential from the solution. An alternative would be to acquire these spectra from a grounded arrangement and under conditions which suppress any extrinsic fields originating from the liquid jet. But without the application of a bias voltage to the solution, $E_{cut}$ cannot be distinguished from the overlapping $E_{cut,HEA}$ of the electron analyzer. In conclusion, there is currently no feasible method to unequivocally determine eΦ changes from aqueous solutions of arbitrary concentration; a detailed discussion is found in our recent report.[1] The exact origin of the observed energy shifts (change of $VIE_{1b1,TBAI}$) thus remains unresolved, and arguably cannot be answered with the currently available experimental tools. To complicate things further, rigid spectral shifts are very common for semiconductors, arising from a local imbalance of charge near the surface which leads to the build-up of a local field.[48, 52-54] Specifically, in the present case, dissolution of salt in water produces hydrated anions and cations which can be viewed as ionized dopants freely moving in the aqueous solution. Charge transfer to the surface leads to a band bending (BB) within a space-charge layer of typically several tens of nm thickness depending on the doping level, manifesting in a rigid spectral energy shift. In the present case, BB is argued to be induced in response to TBAI surface aggregation, which changes the charge distribution at the liquid-vacuum interfacial layer. Arguably, we observe an upward BB, *i.e.*, in the direction of lower VIEs, which is caused by depletion of the solvent's electron density near the surface. The hydrophobic TBA$^+$ molecules which reside near the solution's surface are thought to draw I$^-$ ions into this surface region.[34] It can then be argued that the solvation of I$^-$ reduces water's local electronic density, leading to the observed effect. Notably, the Fermi level remains fixed, or is pinned, within the solution at its bulk value, and aligned with the analyzer; for more details we refer to Ref. [1]. Notably, there would be a rather straightforward experimental test – at least conceptually – to confirm BB. Specifically, illuminating the liquid jet with photons of energy higher than the band gap would generate electron-hole pairs which separate in the electric field of the space-charge layer. This would partially compensate the band bending and induce a surface photovoltage (SPV). In a two-color pump-probe PES experiment one would thus generate a transient flat band, corresponding to the magnitude of the SPV. Currently, one of our labs is being equipped with a VUV source that would in principle allow such an experiment to be performed.

## IV. Conclusions

We have reported a first PES study that quantifies the absolute energetics of aqueous solution ionization as a function of solute concentration. Specifically, lowest vertical ionization energies, VIE, of the water solvent and iodide solute, exemplified for NaI and the surface-active tetrabutylammonium iodide (TBAI) salts, were measured from a liquid microjet using a 40.814 eV photon energy. Our study is a consequent extension of our most recent work that introduced an advanced liquid-jet PES method,[1] based on the measurement of the spectral low-energy cutoff, enabling the determination of absolute ionization energies of solute and solvent. The novelty is that with this more powerful method, previous unsatisfactory gas-phase energy referencing is no longer required. Furthermore, the advanced method enables access to explicit surface and interfacial properties of liquid water and aqueous solutions. For NaI aqueous solution the measured concentration-dependent lowest-ionization energies vary only slightly, up to +260 meV towards larger binding energies in going from dilute to near-saturated solutions. This is largely attributed to associated changes of the bulk-solution electronic structure. The results can be explained with existing theoretical simulations. TBAI, a strong surfactant, exhibits an overall very different behavior, however. Here, VIEs vary to a much greater degree, up to 0.7 eV towards lower binding energies, upon formation of a complete TBAI surface aggregation layer. Such large changes cannot be attributed solely to a change of solute and water electronic structure within the surface monolayer. We provide evidence, supported by a simple estimate of molecular surface-dipole density and orientation and our previous work,[1] that work-function changes play a crucial role. However, we cannot yet rule out contributions of band bending to the observed shifts. To our knowledge, the latter aspect has not been considered in any previous study, other than our own,[1] and shows the importance of exploring such effects both experimentally and theoretically in the future. In a broader context, the present work demonstrates an example of a systematic study quantifying solute- and concentration-dependent absolute electronic energetic changes in aqueous solutions. Application of the new method to other solutions, aqueous or otherwise, is correspondingly straight-forward.

## Data Availability

The data of relevance to this study have been deposited at the following DOI: 10.5281/zenodo.5283376

## Conflict of interest

There are no conflicts to declare.

## Acknowledgements

B. C. acknowledges funding by the SNF (project 200021E-171721) and the EPFL-MPG doctoral school. B. C., S. M., U. H., and B. W. acknowledge support by the Deutsche Forschungsgemeinschaft (Wi 1327/5-1). F. T. and B. W. acknowledge support by the MaxWater initiative of the Max-Planck-Gesellschaft. B. W. acknowledges funding from the European Research Council (ERC) under the European Union's Horizon 2020 research and investigation programme (grant agreement No. 883759). S. T. acknowledges support from the JSPS KAKENHI Grant No. JP20K15229.

# Figures

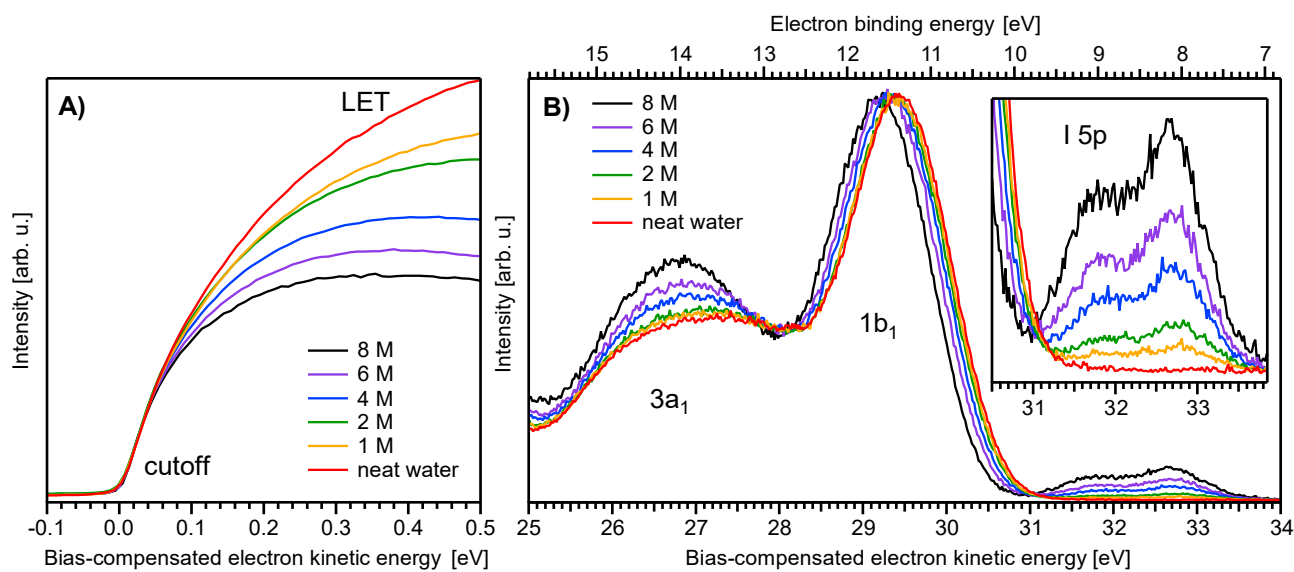

**Figure 1:** Series of experimental spectra for NaI aqueous-solutions of varying salt concentration, spanning neat water (50 mM salt added only for the purpose of maintaining conductivity) to 8 M. All spectra have been energy-shifted to yield $E_{cut} = 0$ eV after applying the tangent-method, *i.e.*, the bottom energy scale shows the KE of the electrons with just enough energy to traverse the liquid surface. **A)** Low-energy tail (LET) spectra with the characteristic cutoff; spectra have been normalized to produce the same tangent slope. An overview of the changes in the wide-range LET shape is shown in Fig. 4. **B)** Valence region with the prominent water $3a_1$ and $1b_1$ bands; spectra have been normalized to the same height of the $1b_1$ peak for visualizing the subtle shifts of the $1b_1$ peak and the shape change of the $3a_1$ peak with increasing concentration. The inset shows an enlarged view of the I⁻ 5p lowest ionization features of the solute. As-measured spectra are plotted in Fig. SI-1 in the Supporting Information.

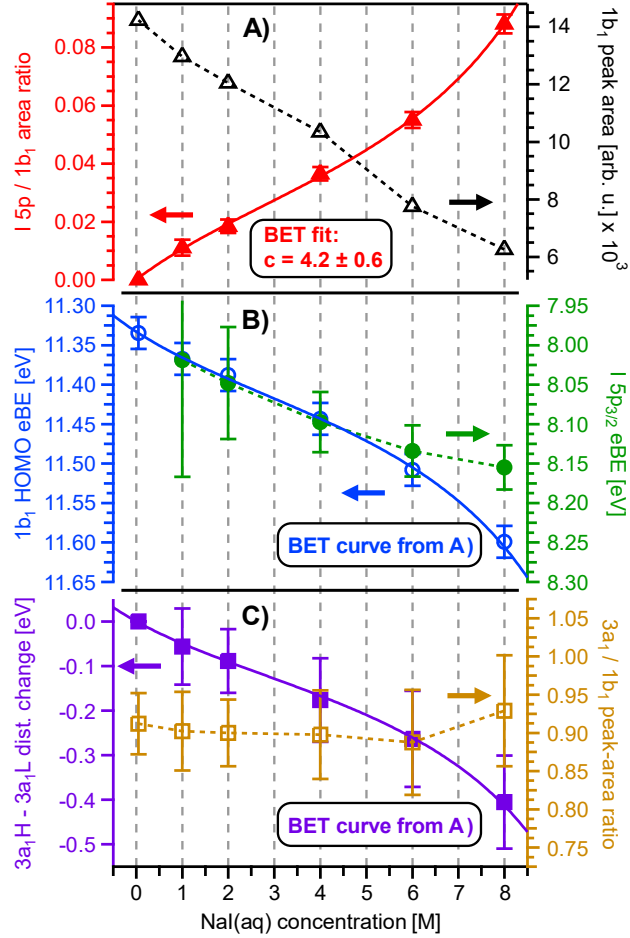

**Figure 2:** Results for NaI$_{(aq)}$ solutions extracted from fits to the spectra plotted as a function of salt concentration (bottom axis). **A)** Solute I⁻ 5p peak area normalized by the 1b$_{1(l)}$ peak area in red (full triangles; left axis) and absolute 1b$_{1(l)}$ peak area in black (open triangles; right axis). The I⁻ 5p peak successively increases in relative intensity while the liquid-water features (represented by the 1b$_1$ intensity) diminish due to reduced relative concentration and enhanced scattering in the surface layer. No saturation behavior is observed for the NaI solute, and instead the trend rather steepens at concentrations above 4 M. A BET (Brunauer, Emmett, and Teller) isotherm was fitted to the data (red line), which yields an excellent agreement with the experimental results (see text for details). **B)** Electron binding energy (eBE) of water's 1b$_{1(l)}$ peak in blue (open circles; left axis) and the I⁻ 5p peak in green (full circles; right axis). Both features shift slightly towards higher eBEs by the same amount but deviate somewhat towards very high concentrations. The 1b$_{1(l)}$ peak eBE follows the surface enrichment of I⁻ 5p 1:1, which is apparent from the excellent match to the BET curve (reproduced here as blue curve by shifting and scaling the red fit curve from panel A). In case of the saturation-like behavior of the I⁻ 5p peak, it can be assumed that the large surface enrichment above 4 M concentration significantly diminishes the solvation of I⁻, which partly compensates the increase in eBE. **C)** Change in energetic splitting of the 3a$_1$ double peak in purple (full squares; left axis) and 3a$_1$ / 1b$_1$ peak-area ratio in orange (open squares; right axis). The overall peak splitting decreases rapidly with increasing concentration while the peak-area ratio stays constant, *i.e.*, the 3a$_1$ feature only seems to increase in relative height because of the diminishing peak distance. Again, the BET curve was reproduced in purple for comparison. The observed narrowing of the 3a$_1$-peak split is in excellent agreement with the values of Ref. [8].

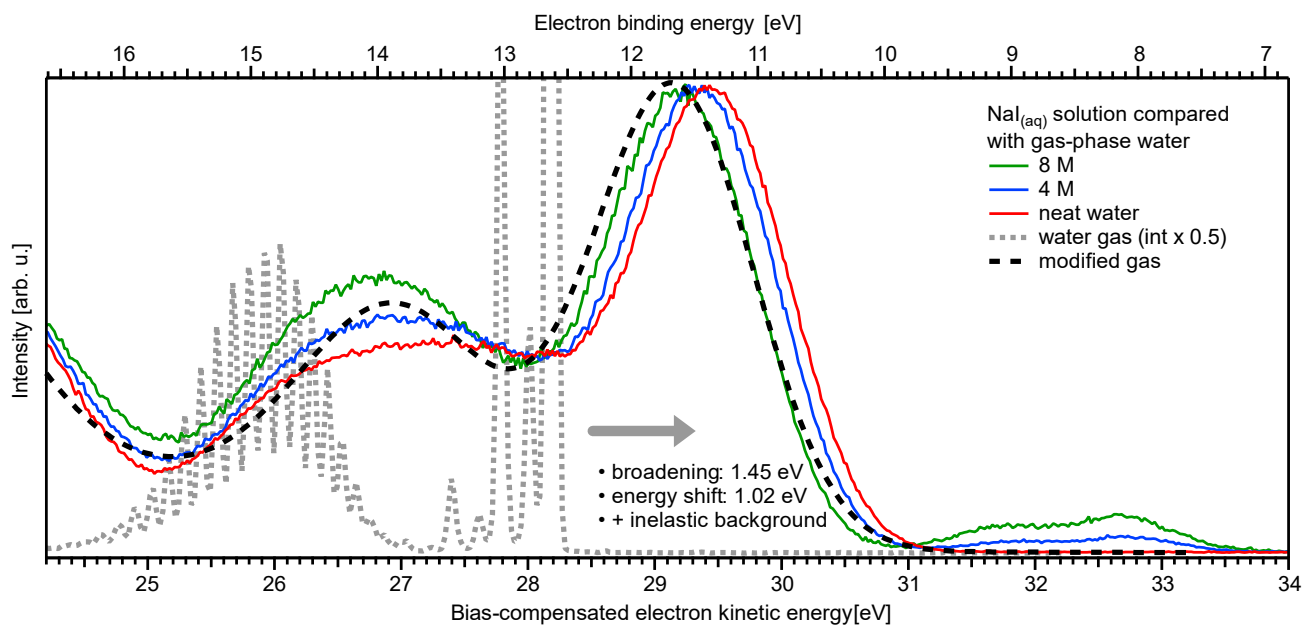

**Figure 3:** Selected spectra from Fig. 1B for neat water (red), and concentrations of 4 M (blue) as well as 8 M (green) of NaI$_{(aq)}$ in comparison with water gas-phase spectra. A high-resolution gas-phase spectrum is plotted as gray dotted line. Some modifications are applied to this spectrum to yield the spectrum plotted as black dashed line: The gas-phase spectrum was convoluted with a Gaussian of FWHM = 1.45 eV, in accordance with the liquid 1b$_1$ peak width reported in Ref. [4], and shifted by 1.02 eV to higher electron KEs (lower BEs), which corresponds to the gas-liquid shift of 1.28 eV (12.62 eV-11.34 eV)[1] for neat liquid water corrected by the 0.26-eV shift after adding 8 M NaI. This modification simulates the unspecific configuration interaction and polarization screening inside the liquid environment. Furthermore, a simple Shirley-type background has been added to include the effect of inelastic scattering for better comparability. The measured 8 M NaI$_{(aq)}$ spectrum (green trace) and the transformed gas-phase spectrum show excellent agreement. Note that any hydrogen-specific effects are absent in the latter, which hints at strongly reduced hydrogen bonding in the 8 M solution.

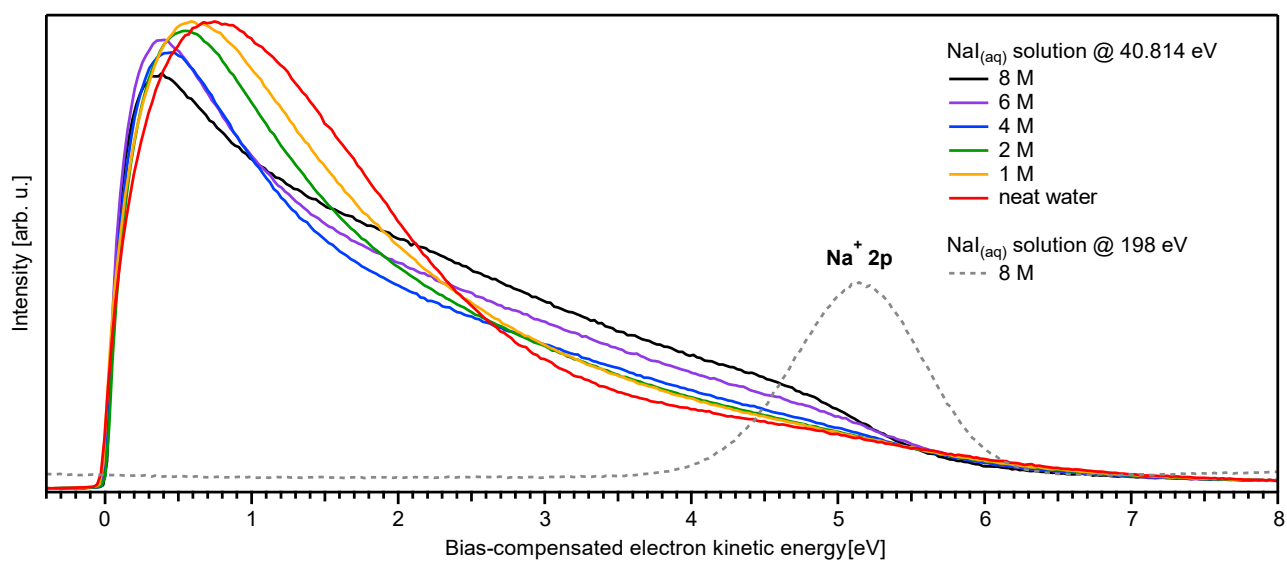

**Figure 4:** Wide-range measurement of the LET for different concentrations of NaI aqueous solution; spectra were normalized to the same scaling factor as in Fig. 1B, *i.e.*, to yield the same height for the 1b$_{1(l)}$ peak feature

(not visible here). A pronounced shape change is observed with increasing salt concentration, especially in the 1-5 eV region. Comparison with data of 8 M NaI$_{(aq)}$ measured at 198 eV (from our previous study)[8] reveals the origin of this signal: The intense Na$^+$ 2p solute feature would appear at ~5.1 eV for the implemented photon energy of 40.814 eV. However, this is already below the critical energy limit of ~10-13 eV to observe undisturbed peak features in liquid water, as recently reported in Ref. [2], and electrons at lower electron KE are subject to strong inelastic scattering, which heavily distorts and diminishes the Na$^+$ peak observed here.

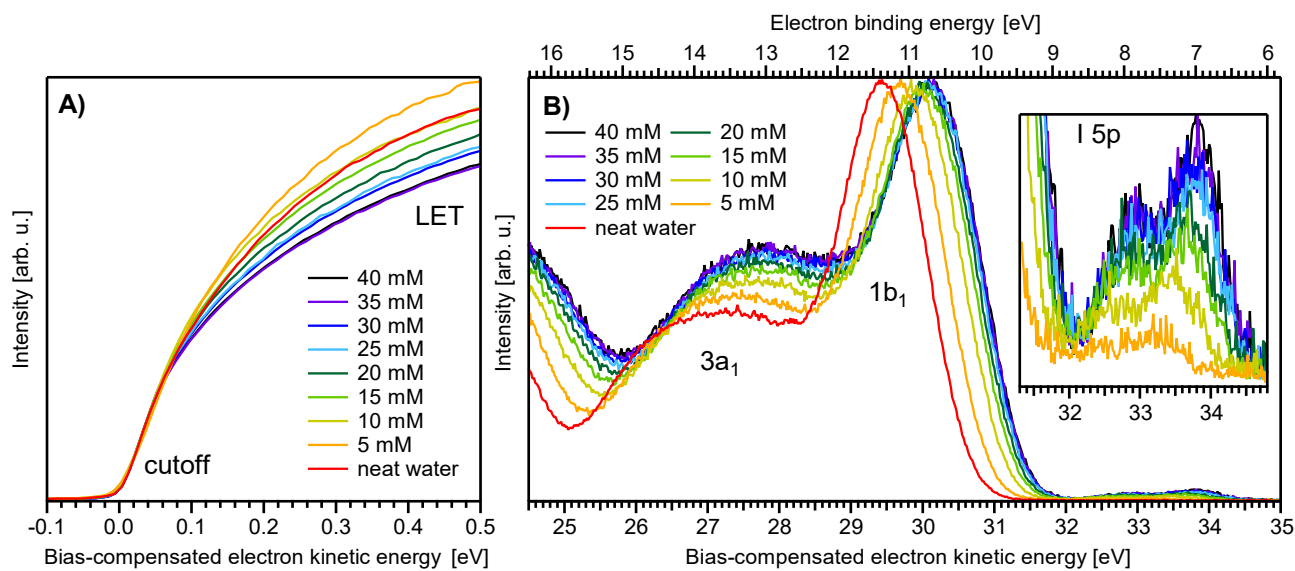

**Figure 5:** Series of TBAI aqueous-solution spectra spanning neat water (with 50 mM NaI added only for the purpose of maintaining conductivity) to 40 mM surface-active salt concentrations in 5 mM steps. The energy scale of all spectra has been shifted to yield E$_{cut}$ = 0 eV after applying the tangent method, *i.e.*, the bottom energy scale shows the kinetic energy of the electrons just after leaving the liquid surface. **A)** Low-energy tail (LET) spectra with the characteristic cutoff; spectra have been normalized to the same tangent slope. An overview of changes in the wide-range LET-shape is shown in Fig. 7. **B)** Valence region with the prominent water 3a$_1$ and 1b$_1$ bands; spectra have been normalized to the same height of the 1b$_1$ peak for visualizing the 1b$_1$ peak shifts and 3a$_1$ peak-shape changes with increasing concentration. The saturation behavior where the spectra converge to a final form is apparent. The inset shows an enlarged view on the I$^-$ 5p lowest ionization feature of the solute. Fig. SI-2 of the Supporting Information shows the as-measured spectra.

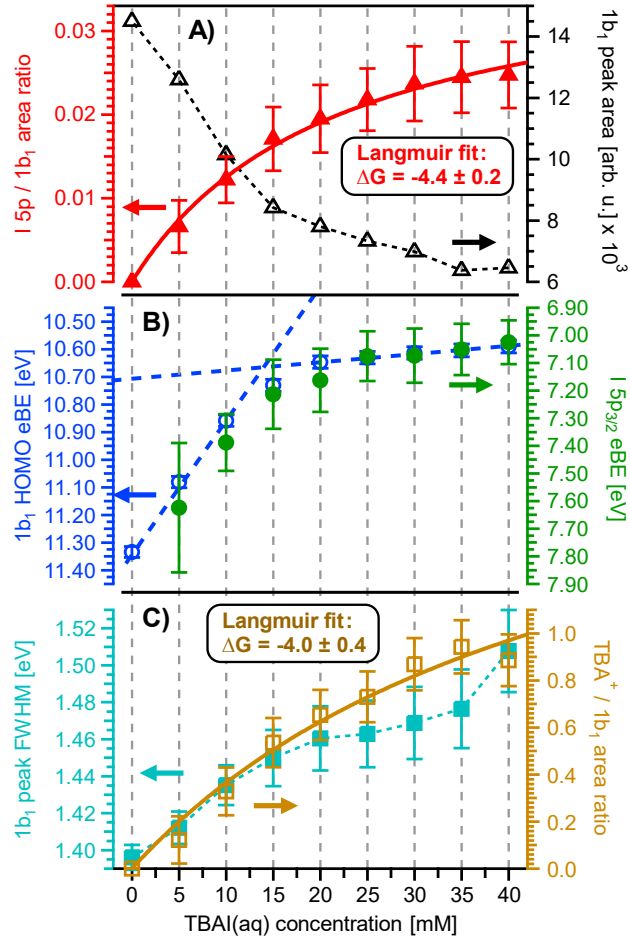

**Figure 6:** Results for TBAI$_{(aq)}$ solutions extracted from fits to the spectra plotted as a function of salt concentration (bottom axis), similar to Fig. 2. **A)** Solute I$^-$ 5p peak area normalized by the 1b$_1$ peak area in red (full triangles; left axis) and absolute 1b$_1$ peak area in black (open triangles; right axis). The I$^-$ 5p peak successively increases in relative intensity, while the liquid-water features (represented by the 1b$_1$ intensity) diminish due to enhanced scattering in the surface layer. Saturation behavior is observed for both signals above 20 mM. The I$^-$ 5p peak-area data has been fitted to a Langmuir adsorption isotherm (red line; see text for detail). **B)** Electron binding energy (eBE) of water's 1b$_{1(l)}$ peak in blue (open circles; left axis) and the I$^-$ 5p peak in green (full circles; right axis). Both features shift rapidly towards lower eBEs by the same amount. A steep decrease is observed at lower concentrations, coinciding with the filling of the first monolayer, and then increases only slowly afterwards (blue dashed lines added as a guide to the eye). **C)** Change in 1b$_1$ peak width in cyan (full squares; left axis) and the TBA$^+$ / 1b$_1$ peak-area ratio in orange (open squares; right axis). Here, the TBA$^+$ signal is taken from the difference spectra between neat water and various concentrations of TBAI$_{(aq)}$, the difference spectra are plotted in Fig. 7. The normalized TBA$^+$ feature increases in intensity similar to the I$^-$ 5p peak, this data has been fitted to a Langmuir curve as well (orange line). It is inferred that all water PE features get broader with increasing solute concentration, which is exemplified by the increasing 1b$_1$ peak FWHM. The width increase of all features in the spectrum may originate from altered scattering behavior on the surface layer of the solution or an increase in the hydration configurations sample as the interfacial concentration is increased. The evolution in shape of the valence spectra is shown in Fig. SI-4 of the Supporting Information.

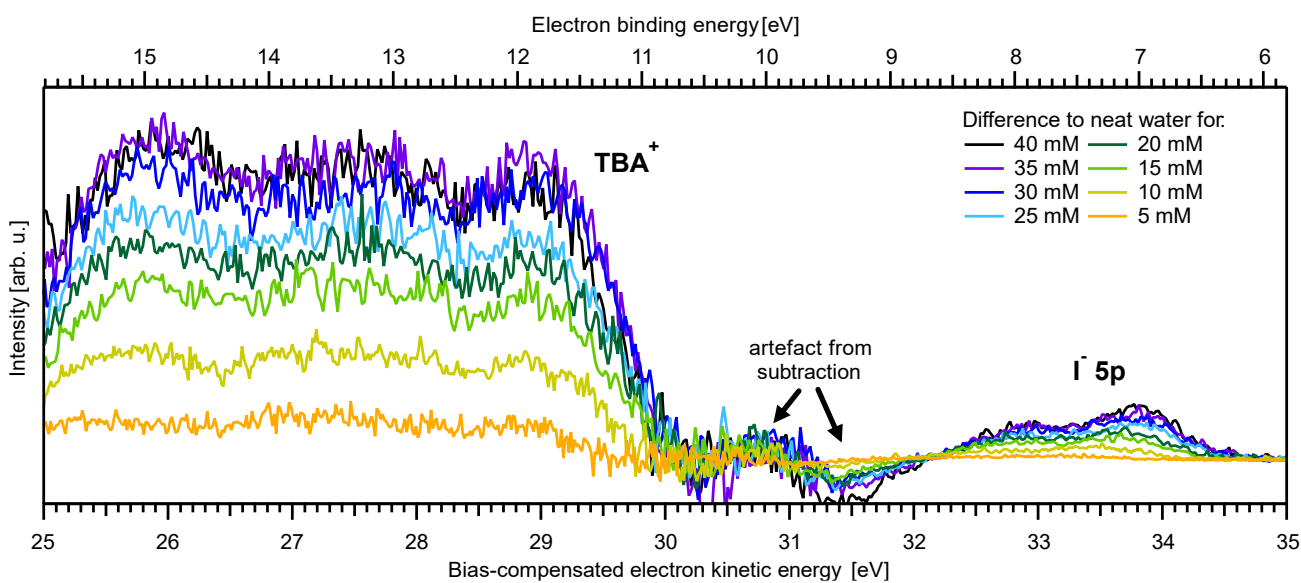

**Figure 7:** Difference between the neat water spectrum and spectra for various concentrations of TBAI$_{(aq)}$ after normalization to the same 1b$_1$ peak height, *i.e.*, it is assumed that solute contributions below the 1b$_1$ peak are zero. The reference spectrum (not shown, see red curve in Figs. 5B and SI-4) has also been successively Gaussian-broadened before calculating the difference to account for the broadening effect observed with higher TBAI concentration (compare to the 1b$_1$ FWHM in Table 2). The signal contribution at an eKE of 25-30 eV is assigned to TBA$^+$ and increases in intensity similarly to the I$^-$ 5p signal at 32-35 eV (see Fig. 6C).

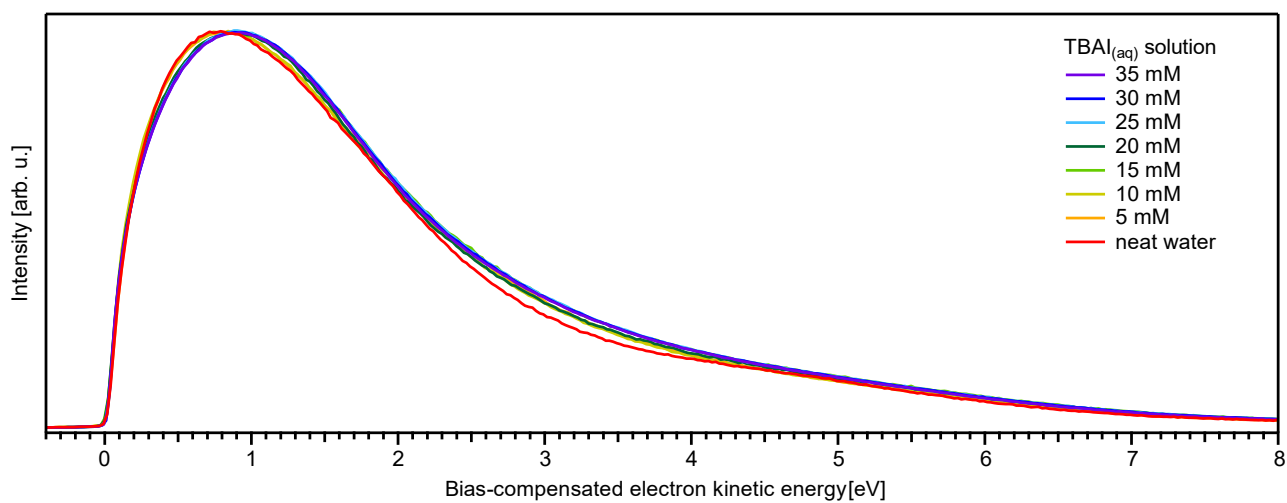

**Figure 8:** Wide-range measurement of the LET for different concentrations of TBAI$_{(aq)}$ normalized to the same maximum height. Only slight changes in LET shape are observed for TBAI$_{(aq)}$. Most notable is a slight signal rise near ~3 eV KE which can be crudely attributed to the corresponding inelastic scattering maximum of the TBA$^+$ features. This feature increases in intensity in a similar way to the primary TBAI$^+$ photoelectron peaks and is found at approximately ~24 eV higher KE, where 20-25 eV energy loss corresponds to the maximum in the inelastic scattering probability for water.[42, 43]

# TOC Graphic

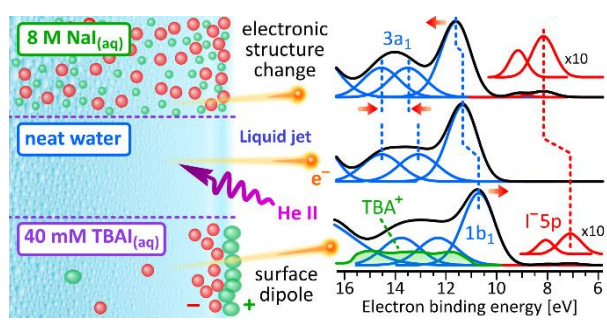

**Caption:** Significant differences are observed in liquid-water's lowest electron binding energy with increasing solute concentration in archetypal aqueous solutions. For $NaI_{(aq)}$ and $TBAI_{(aq)}$, the energy changes extend to +0.3 eV and -0.7 eV, respectively.

# Tables

| conc. | $VIE_{1b1}$ (eV) | $VIE_{3a1L}$ (eV) | $VIE_{3a1H}$ (eV) | $VIE_{I5p1/2}$ (eV) | $VIE_{I5p3/2}$ (eV) | $3a_1$ split (eV) |
|---|---|---|---|---|---|---|
| 50 mM | 11.33 ± 0.02 | 13.09 ± 0.05 | 14.53 ± 0.05 | - | - | - |
| 1 M | 11.37 ± 0.02 | 13.14 ± 0.06 | 14.54 ± 0.06 | 9.00 ± 0.13 | 8.02 ± 0.15 | -0.06 ± 0.09 |
| 2 M | 11.39 ± 0.02 | 13.17 ± 0.05 | 14.53 ± 0.05 | 9.03 ± 0.07 | 8.05 ± 0.07 | -0.09 ± 0.07 |
| 4 M | 11.44 ± 0.02 | 13.26 ± 0.07 | 14.53 ± 0.07 | 9.06 ± 0.04 | 8.10 ± 0.03 | -0.18 ± 0.09 |
| 6 M | 11.51 ± 0.02 | 13.35 ± 0.07 | 14.53 ± 0.08 | 9.12 ± 0.03 | 8.14 ± 0.03 | -0.26 ± 0.11 |
| 8 M | 11.60 ± 0.02 | 13.50 ± 0.07 | 14.55 ± 0.07 | 9.16 ± 0.02 | 8.16 ± 0.02 | -0.41 ± 0.10 |

**Table 1:** VIE values of the liquid water valence $1b_1$ and split $3a_1$ bands (denoted as $3a_1$ H and $3a_1$ L; see text) as well as the solute $I^-$ 5p doublet peak as extracted from fits to the spectra of solutions with various NaI concentrations. The right-most column shows the change in energetic distance between the $3a_1$ H and $3a_1$ L bands, which increases with increasing NaI concentration. Errors are one standard deviation as derived from the fits.

| conc. | $VIE_{1b1}$ (eV) | $VIE_{3a1L}$ (eV) | $VIE_{3a1H}$ (eV) | $VIE_{I5p1/2}$ (eV) | $VIE_{I5p3/2}$ (eV) | $1b_1$ FWHM (eV) |
|---|---|---|---|---|---|---|
| 0 mM | -11.33 ± 0.02 | 13.12 ± 0.03 | 14.52 ± 0.03 | - | - | 1.40 ± 0.01 |
| 5 mM | -11.08 ± 0.02 | 12.85 ± 0.03 | 14.25 ± 0.04 | 8.63 ± 0.25 | 7.63 ± 0.23 | 1.41 ± 0.01 |
| 10 mM | -10.86 ± 0.02 | 12.61 ± 0.04 | 14.01 ± 0.04 | 8.35 ± 0.12 | 7.39 ± 0.10 | 1.44 ± 0.01 |
| 15 mM | -10.73 ± 0.02 | 12.47 ± 0.05 | 13.87 ± 0.05 | 8.19 ± 0.12 | 7.21 ± 0.12 | 1.45 ± 0.02 |
| 20 mM | -10.65 ± 0.02 | 12.40 ± 0.06 | 13.80 ± 0.06 | 8.12 ± 0.11 | 7.16 ± 0.11 | 1.46 ± 0.02 |
| 25 mM | -10.63 ± 0.02 | 12.33 ± 0.05 | 13.73 ± 0.05 | 8.05 ± 0.10 | 7.08 ± 0.09 | 1.46 ± 0.02 |
| 30 mM | -10.61 ± 0.02 | 12.29 ± 0.05 | 13.69 ± 0.05 | 8.03 ± 0.12 | 7.08 ± 0.10 | 1.47 ± 0.02 |
| 35 mM | -10.60 ± 0.02 | 12.25 ± 0.06 | 13.65 ± 0.05 | 8.02 ± 0.10 | 7.05 ± 0.09 | 1.48 ± 0.02 |
| 40 mM | -10.59 ± 0.02 | 12.34 ± 0.05 | 13.74 ± 0.05 | 7.99 ± 0.10 | 7.03 ± 0.08 | 1.51 ± 0.02 |

**Table 2:** VIE values of the liquid water valence $1b_1$ and split $3a_1$ bands ($3a_1$ H and $3a_1$ L; see text) as well as the solute $I^-$ 5p doublet peak as extracted from fits to the spectra of solutions with various TBAI concentrations. The right-most column shows the full-width at half maximum (FWHM) of the $1b_1$ band, which increases with increasing TBAI concentration; this is deemed to be representative of an overall broadening of all water bands. Errors are one standard deviation as derived from the fit.

## Notes and references

N1  Note that alternative explanations for a plateau feature of the intensity observed in the LET curve region shown in Figure 4 – such as electron-impact-induced electronic transitions to the charge-transfer-to-solvent (CTTS) bands in the $I^-_{(aq)}$ anion[55] – can be readily excluded as primary origins of these features. As shown in Figures 4 and 8 and further discussed below, the ~3 eV plateau features are absent from the neat water spectra and the most concentrated TBAI$_{(aq)}$ solution spectra, which correspond to surface iodide concentrations locally equivalent to 3 M bulk iodide concentrations. Consequently, the plateau feature in Figure 4 can be directly related to the Na$^+_{(aq)}$ cation. We further note that there is no evidence for a 5-7 eV electron KE loss channel associated with the dominant, directly emitted water 1b1 photoelectron peak.

# Supporting Information

## Quantitative electronic structure and work-function changes of liquid water induced by solute

Bruno Credidio[1,2], Michele Pugini[1], Sebastian Malerz[1], Florian Trinter[1,3], Uwe Hergenhahn[1], Iain Wilkinson,[4] Stephan Thürmer[5]*, and Bernd Winter[1]*

[1] *Molecular Physics Department, Fritz-Haber-Institut der Max-Planck-Gesellschaft, Faradayweg 4-6, 14195 Berlin, Germany*
[2] *Institute for Chemical Sciences and Engineering (ISIC), École Polytechnique Fédérale de Lausanne (EPFL), 1015 Lausanne, Switzerland*
[3] *Institut für Kernphysik, Goethe-Universität, Max-von-Laue-Straße 1, 60438 Frankfurt am Main, Germany*
[4] *Department of Locally-Sensitive & Time-Resolved Spectroscopy, Helmholtz-Zentrum Berlin für Materialien und Energie, Hahn-Meitner-Platz 1, 14109 Berlin, Germany*
[5] *Department of Chemistry, Graduate School of Science, Kyoto University, Kitashirakawa-Oiwakecho, Sakyo-Ku, Kyoto 606-8502, Japan*

# Figures

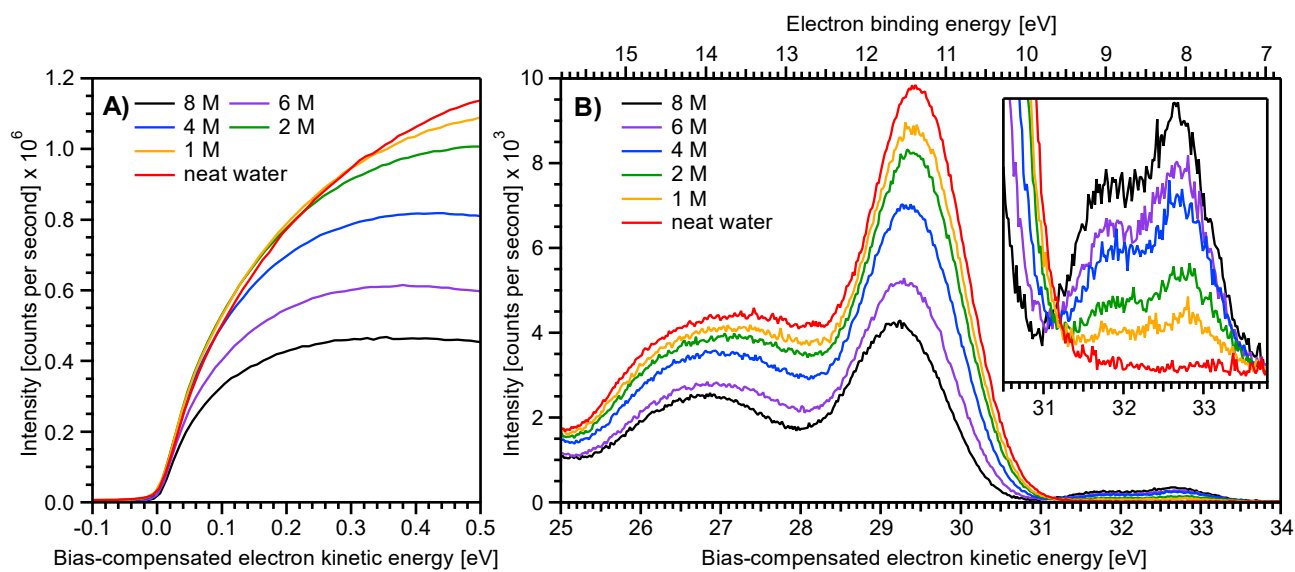

**Figure SI-1:** The same data for NaI$_{(aq)}$ as in Fig. 1, but here the intensity is shown as measured. The water signal decreases with higher NaI concentration.

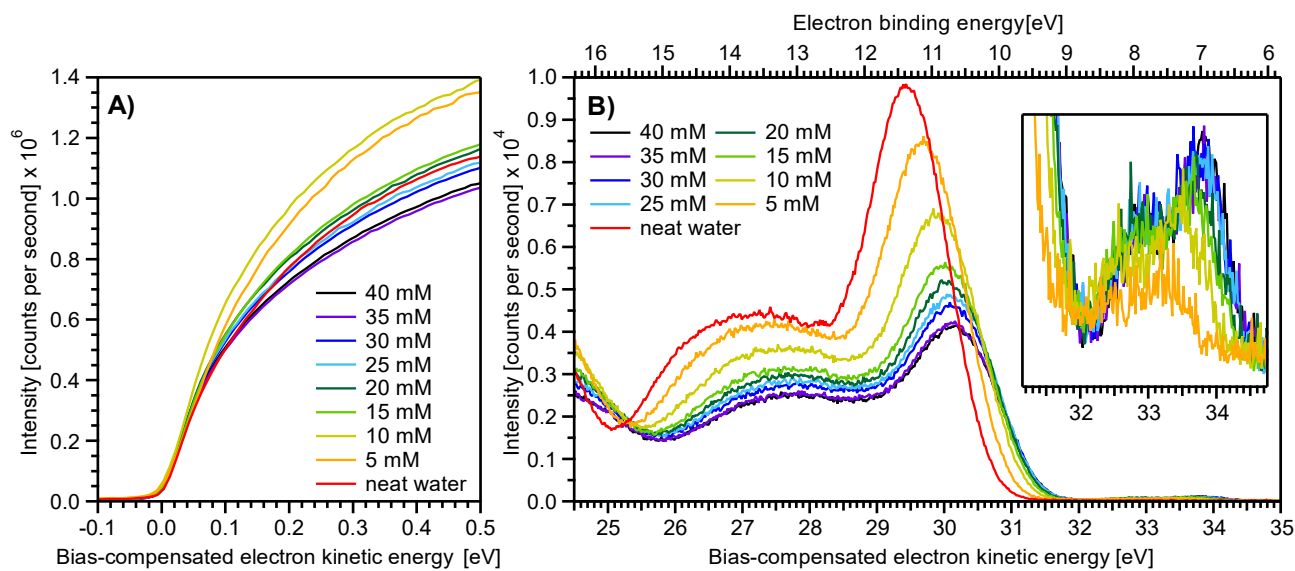

**Figure SI-2:** The same data for TBAI$_{(aq)}$ as shown in Fig. 5, but here the intensity is shown as measured. Similar to NaI, the water signal decreases with higher TBAI concentration.

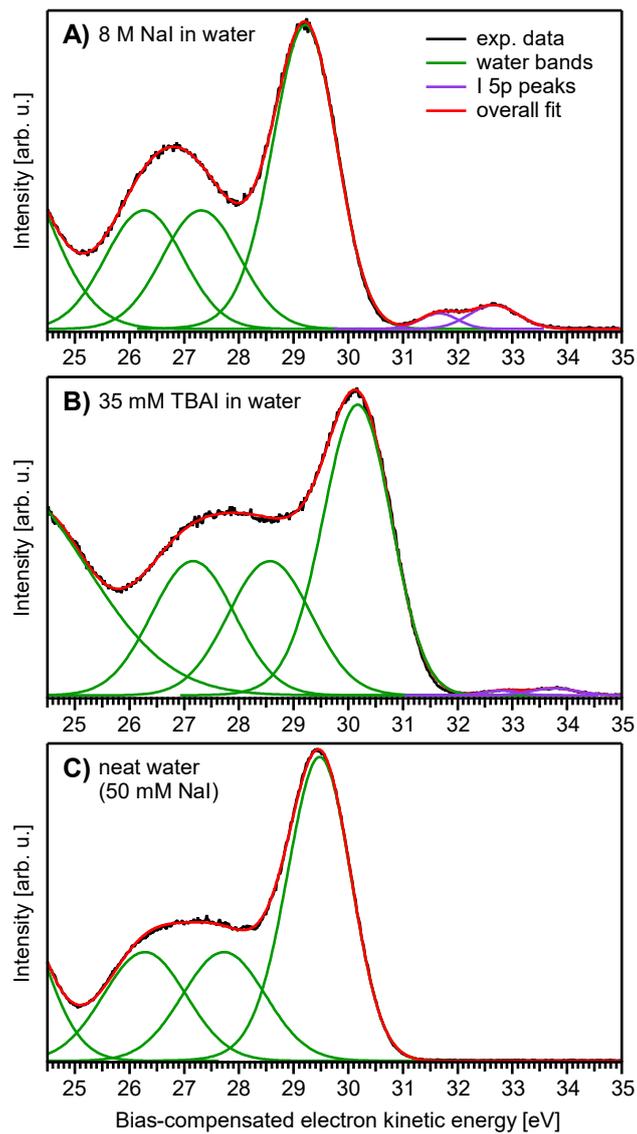

**Figure SI-3:** Exemplary fits to the spectra: **A)** 8 M NaI solution, **B)** 35 mM TBAI solution, and **C)** neat water (*i.e.*, with only 50 mM NaI added for charge compensation and to enable sample biasing); measured data in black, the overall fit in red, water-band features in green, and the I⁻ 5p doublet peak in violet. See text for details.

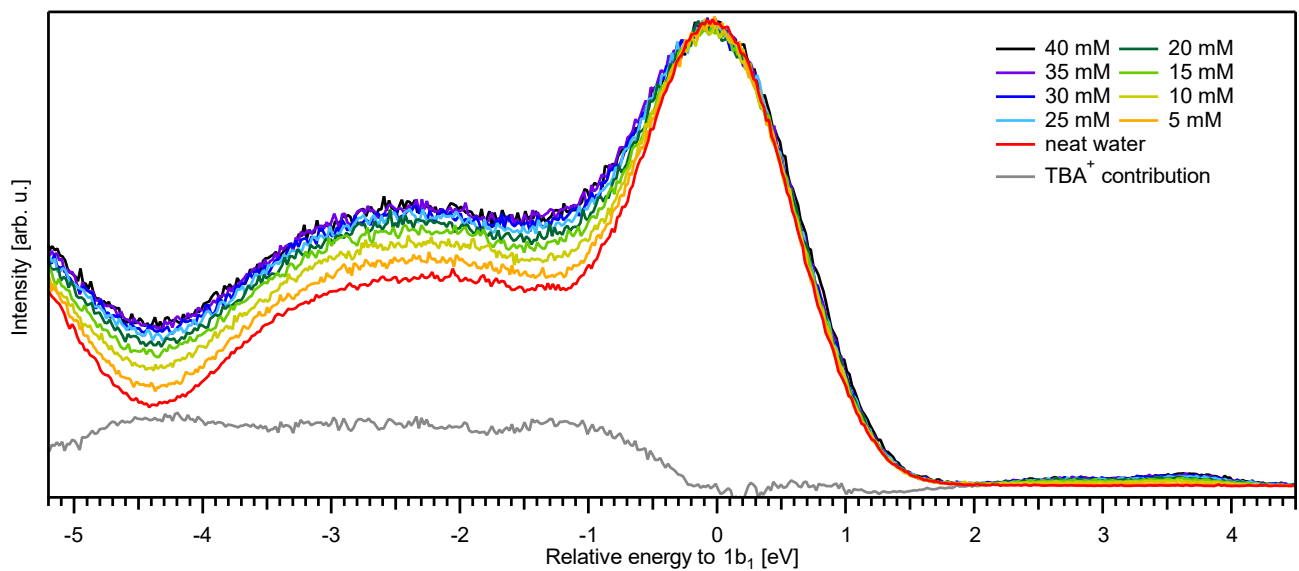

**Figure SI-4:** The same data for TBAI$_{(aq)}$ as in Figs. 5B and SI-2B but aligned to the same 1b$_1$ peak position for better comparison of spectral changes with increasing concentration. The grey curve shows the difference between the 35-mM TBAI$_{(aq)}$ and neat water spectrum, which we assign to TBA$^+$.